\documentclass[%
 reprint,
superscriptaddress,
 amsmath,amssymb,
 aps,
]{revtex4-1}

\usepackage{graphicx}
\usepackage{dcolumn}
\usepackage{bm}

\begin{document}

\preprint{APS/123-QED}

\title{Terahertz-frequency magnon-phonon-polaritons in the strong coupling regime}

\author{P. Sivarajah}
\author{J. Lu}
\affiliation{%
Massachusetts Institute of Technology, Cambridge, MA 02139, USA}%
\author{M. Xiang}
\author{W. Ren}
\affiliation{%
Department of Physics, International Center of Quantum and Molecular Structures, and Materials \\ Genome Institute, Shanghai University, Shanghai 200444, China
}%

\author{S. Kamba}
\affiliation{
Institute of Physics, Academy of Sciences of the Czech Republic, Na Slovance 2, 182 21 Prague 8, Czech \\ Republic
}%

\author{S. Cao}
 \email{sxcao@shu.edu.cn}
\affiliation{%
Department of Physics, International Center of Quantum and Molecular Structures, and Materials \\ Genome Institute, Shanghai University, Shanghai 200444, China
}%

\author{K. A. Nelson}
\affiliation{%
Department of Chemistry, Massachusetts Institute of Technology, Cambridge, MA 02139, USA}%

\date{\today}

\begin{abstract}
Strong coupling between light and matter occurs when the two interact strongly enough to form new hybrid modes called polaritons. Here we report on the strong coupling of both the electric and magnetic degrees of freedom to an ultrafast terahertz (THz) frequency electromagnetic wave. In our system, optical phonons in a slab of ferroelectric lithium niobate (LiNbO$_3$) are strongly coupled to a THz electric field to form phonon-polaritons, which are simultaneously strongly coupled to magnons in an adjacent slab of canted antiferromagnetic erbium orthoferrite (ErFeO$_3$) via the THz magnetic field. The strong coupling leads to the formation of new magnon-phonon-polariton modes, which we experimentally observe in the wavevector-frequency dispersion curve as an avoided crossing, and in the time-domain as a normal-mode beating. Our simple yet versatile waveguide provides a promising avenue by which to explore ultrafast THz spintronics applications. 
\end{abstract}

\pacs{Valid PACS appear here}
\maketitle


\section{\label{sec:Introduction}Introduction}

The interaction between an electromagnetic wave (light) and an atomic, molecular, or material mode is termed “strong coupling” if the rate of coherent energy transfer between the light and matter is faster than the irreversible decay of the light or the coherent material excitation. In the strong coupling regime, the electromagnetic and material modes can no longer be treated as separate entities but rather form two hybrid modes called polaritons, with an energy difference between the modes given by the splitting energy $\hbar\Omega$~\cite{Kwek2014,Mills2001,Snoke2015,Torma2015}. Polariton systems enable extensive optical control over material behavior and coherent information transfer between light and material degrees of freedom at a rate $\Omega$, yielding opportunities for both classical~\cite{Aspelmeyer2014,Barnes2003} and quantum information processing~\cite{Imamoglu1999,Monroe2002,You2003}. Early on, the strong coupling of the electric field of light to the electric dipole moments of atoms was explored~\cite{Thompson1992}, and later on was also demonstrated with superconducting qubits~\cite{Wallraff2004} and excitons~\cite{Bellessa2004,Kasprzak2006a,Reithmaier2004,Yoshie2004}. More recently, coupling to magnetic moments has been studied in the microwave domain using magnetically active systems such as NV centers~\cite{Kubo2010} and magnon spin waves~\cite{Tabuchi2014,Zhang2014} to form a foundation for the next generation in spintronics~\cite{Awschalom2013}. Largely, the focus of strong coupling physics has been on either the electric or the magnetic degree of freedom. However, the ensuing physics and potential applications motivate the prospect of coupling both electric and magnetic dipoles to an electromagnetic wave, while maintaining the ability to individually tailor and address each. Spintronics, in its quest toward long-range and terahertz (THz) frequency operation, would particularly benefit from such strong coupling because it provides a means for facile transport and interaction with spin information. In recent advances of microwave spintronics, magnons have been used as data carriers~\cite{Chumak2012,Kajiwara2010}, because, unlike spin-polarized electrons, they provide transfer of spin information over macroscopic distances without any Joule heating and enable access to wave-based computing concepts. However, a similar use of magnons at THz frequencies is difficult due to a scarcity of appropriate electrically based sources and detectors. Optical light can be used for generation and detection of THz magnons, but the interaction is either indirect~\cite{Ivanov2014,Talbayev2008} or nonlinear~\cite{Kalashnikova2007} and the efficiency is not high. More appealingly, free-space THz light has been shown to exert direct linear control over magnons~\cite{Kampfrath2011,Yamaguchi2010a}, but free-space sources are not conducive to realizing miniaturized devices. These limitations inspire exploration of strong coupling of magnons to both light and other material degrees of freedom to enhance functionality. 

In this article, we demonstrate strong coupling of magnetic and electric dipoles in two materials mediated by an ultrafast THz-frequency electromagnetic wave. We show that by simultaneously strongly coupling the optical phonons in a ferroelectric slab and magnons in an antiferromagnet slab to THz electric and magnetic fields, respectively, it is possible to form new hybrid modes termed magnon-phonon-polaritons. Although it is possible to drive a lattice vibration and magnetization in a multiferroic material that exhibits coupling between magnetic and electric order parameters~\cite{Tokura2014}, here we demonstrate strong light-matter coupling with no need for intrinsic multiferroicity. We recently demonstrated this form of light-matter coupling to THz cavity photons in a high-Q ($\sim 1000$), small mode-volume $V = 3.4\times10^-3 \lambda^3 = 0.5(\lambda/n)^3$ hybrid 3D cavity~\cite{Sivarajah2017}. In contrast, in this work the polaritons propagate over macroscopic distances while coherently exchanging energy between the lattice vibrations and magnetization. We demonstrate these features in an on-chip waveguide platform where the polaritons are highly confined, and efficiently generated and detected at ultrafast time scales using an all-optical method. In Sec. \ref{sec:Generation}, we will first describe our experimental setup, which uses an all-optical method. In Sec. \ref{sec:Avoided}, we will then describe evidence for strong coupling in the form of avoided crossings in the experimentally recorded dispersion curves. Here, we will also discuss a coupled oscillator model that shows excellent agreement to the data. In Sec. \ref{sec:Lorentz}, a Lorentz model for strong coupling is then discussed, and comparisons to the coupled oscillator model are discussed. In Sec. \ref{sec:Absorption}, we will then highlight an experiment that draws distinctions between the signature of strong coupling and simple linear absorption. Finally, in Sec. \ref{sec:Exchange}, evidence for coherent energy exchange between the magnetization in ErFeO$_3$ and polarization in LiNbO$_3$ is demonstrated via normal-mode oscillations. In the Appendix \ref{sec:latticeenergy}, we also discuss the importance of including the phonon mode in our calculations, as it carries a significant fraction of the energy. 

\section{Generation and detection of magnon-phonon-polaritons}\label{sec:Generation}

\begin{figure}
\includegraphics[width=86mm]{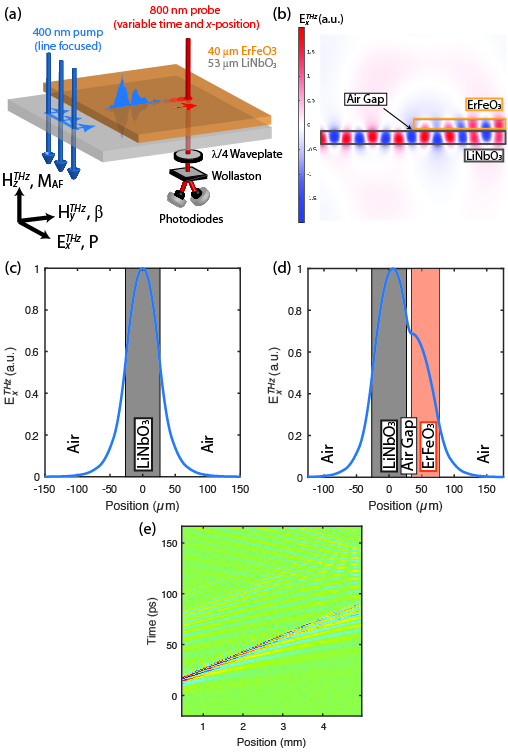}
\caption{\label{fig:setup} Spatiotemporal detection of waveguide THz magnon-phonon-polaritons. (a) A line-focused 400 nm pump pulse generates THz frequency phonon-polaritons in LiNbO$_3$ that enter the LiNbO$_3$--air--ErFeO$_3$ hybrid slab and excite magnons to create magnon-phonon-polaritons. An 800 nm probe pulse is variably delayed to measure the time-dependent THz electric field profile at each of many sample locations in the $y$-direction through the EO effect in LiNbO$_3$. The coordinate axes represent the polarization of the THz light, the waveguide wavevector $\beta$, the ErFeO$_3$ magnetization $M_{AF}$ that is modulated by the AF magnon mode, and the LiNbO$_3$ ferroelectric polarization $P$ that is modulated by the phonon-polariton mode. (b) FEM simulation of electromagnetic wave transmission at 0.67 THz in the absence of the magnon resonance. The colormap of the out-of-plane electric field profile is shown and corresponds to $E_{x}^{THz}$. (c), (d) Calculated $E_{x}^{THz}$ field profile for the first-order waveguide modes in the (c) bare 53 $\mu$m LiNbO3 slab and (d) hybrid 53 $\mu$m LiNbO$_3$--8 $\mu$m air--40 $\mu$m ErFeO$_3$ structure. (e) Space-time plot for the hybrid LiNbO$_3$--air--ErFeO$_3$ slab. At each position along the propagation direction, the time-dependent THz electric field profile is displayed along the vertical axis.}
\end{figure}

Our system consists of a thin composite slab of 53 $\mu$m thick (100) lithium niobate (LiNbO$_3$) and 40 $\mu$m thick (001) erbium orthoferrite (ErFeO$_3$). The LiNbO$_3$ sample was an MgO-doped stoichiometric melt single crystal grown by the Czochralski method and high-quality laser grade polished down to 53 $\mu$m and diced to dimensions of $11 \times 10$ mm. The ErFeO3 sample was a single crystal grown by the floating-zone method. The sample plate with size $5 \times 5$ mm was polished using diamond paste (1 $\mu$m grain size) down to 40 $\mu$m. Large sample planes were prepared plane parallel with accuracy under 1 $\mu$m.  A thin layer of air, approximately 8 $\mu$m thick, separates the two slabs. Note that the optically shiny surfaces had roughness less than 100 nm, whereas the THz radiation has more than one order of magnitude larger wavelength.  In LiNbO$_3$, the material excitation is the polar lattice vibration (i.e. transverse optical phonon mode) with its polarization $P$ along the (ferroelectric) $c$ crystallographic axis of the tetragonal crystal (the $x$-axis as defined in Fig.~\ref{fig:setup}a). The electric field component of a THz-frequency electromagnetic wave, oriented along the $x$-direction, coherently couples to these phonons to produce mixed phonon-polariton modes~\cite{Feurer2007ARM,Feurer2003c}. Our phonon-polariton wavepackets, with $\sim 0.1-2$ THz spectral content, drive the lower polariton branch that extends below the optical phonon frequency of 7.4 THz and corresponds to in-phase-excursions of the lattice vibration and the electromagnetic wave.  Although the polaritons in our experiments are well detuned from the phonon resonance, the phononic nature is still integral to the dynamics. Not only does the lattice carry $\sim 32\%$ of the energy in the 0.1-2 THz range (see Appendix \ref{sec:latticeenergy}), the phonon mode is also responsible mechanistically for efficient generation and detection of the THz waves in our experiments. More specifically, the phonon is responsible for generating the THz waves through an impulsive driving by the incident optical pulse~\cite{Cheung1985}, and allows direct detection of the THz waves through perturbing the optical polarization via electron-lattice interactions~\cite{Kaminow1967}. In ErFeO$_3$, the relevant material excitation is the collective magnetic spin, i.e. the Brillouin zone center quasi-antiferromagnetic (AF) magnon mode wherein the net magnetization $M_{AF}$ along the $c$ axis of the orthorhombic crystal (the $z$-axis as defined in Fig.~\ref{fig:setup}a) is modulated in amplitude~\cite{Shapiro1974,Yamaguchi2010a}. The magnetic field component of THz light along the $z$-axis coherently couples to the AF mode at the magnetic resonance transition frequency of 0.67-THz (with no applied magnetic field) to produce mixed magnon-polaritons. The first report of THz frequency magnon-polaritons was presented Sanders et al. in 1978~\cite{Sanders1978}. The authors used a slab of iron (II) fluoride that exhibits an antiferromagnetic resonance at 1.58 THz to form a strongly coupled with THz light. Here, we complement this interaction by also coupling the THz light to phonons in a highly confined waveguide geometry. In particular, in our hybrid waveguide, uniform $x$-electric-polarized, $z$-magnetic-polarized THz-frequency electromagnetic waves extend throughout both materials (see Fig. \ref{fig:setup}), resulting in new hybridized magnon-phonon-polariton modes. 

To study the interaction, we performed pump-probe measurements using a Ti:Sapphire amplifier system (Coherent Inc.) that outputs 90 fs duration pulses with an 800 nm center wavelength at a 1 kHz repetition rate. The laser power was first split in a ratio of 90/10 for the pump and probe pulses, respectively. The pump pulse was variably delayed by $t$, frequency doubled to 400 nm in a $\beta$-barium borate (BBO) crystal, and as shown in Fig.~\ref{fig:setup}a, cylindrically focused into the bare LiNbO$_3$ portion of the slab to excite a $z$-electric-polarized, $y$-propagating phonon-polariton wavepacket via impulsive stimulated Raman scattering~\cite{Dougherty1992a}. This method has been exploited extensively for THz wave generation in LiNbO$_3$~\cite{Yeh2007a}. As shown in Fig.~\ref{fig:setup}a, the phonon-polaritons generated by the optical pulse propagated away from the generation region and into the hybrid LiNbO$_3$--air--ErFeO$_3$ slab. Here, the electric and magnetic fields of the THz wave, polarized along the $x$- and $z$-directions respectively drive the polarization of the polar phonon mode in LiNbO$_3$ and the magnetization of the AF magnon mode in ErFeO$_3$. In Fig. \ref{fig:setup}b, we show a finite element method (FEM) simulation of electromagnetic wave transmission from the bare slab into the hybrid slab at 0.67 THz (in the absence of any resonances). The transmission can be seen to proceed fairly efficiently. Correspondingly, in Fig. \ref{fig:setup}c and \ref{fig:setup}d we show the FEM calculated field profile for the first-order waveguide modes in bare and hybrid structures at 0.67 THz, respectively. In both the single and hybrid waveguides, a uniform electromagnetic mode can be seen extending through the slab(s) and evanescently decaying away on either side of the structure.

\begin{figure}
\includegraphics[width=86mm]{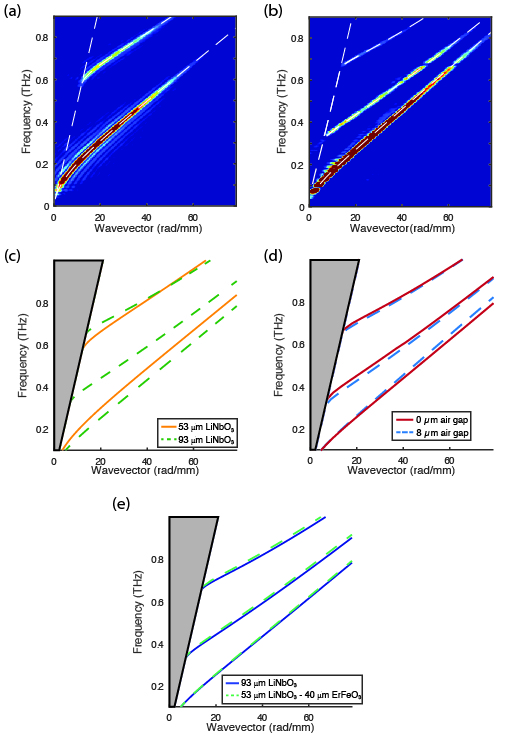}
\caption{\label{fig:dispersion} Dispersion curves. (a) Experimentally recorded TE-waveguide dispersion curves determined from measurements of the bare LiNbO$_3$ slab, showing the first two modes along with a numerical solutions (dashed curves). (b) In the hybrid LiNbO$_3$--air--ErFeO$_3$ slab, a clear signature of an avoided crossing can be seen at the magnon resonance frequency of 0.67 THz in both the first and second-order modes. (c) Calculated dispersion curves in the absence of the magnon resonance for a LiNbO$_3$ slab (i.e. assuming $n=5.0$) of thickness of 53 $\mu$m and 93 $\mu$m. (d) Dispersion curves for a hybrid structure of 53 $\mu$m LiNbO$_3$--8 $\mu$m air (i.e assuming $n=1.0$)--49 $\mu$m ErFeO$_3$ (i.e. assuming $n=4.9$) and 53 $\mu$m LiNbO$_3$--49 $\mu$m ErFeO$_3$ (i.e. without air gap) demonstrate the minimal influence of an air gap. (e) Dispersion curves for a 93 $\mu$m LiNbO$_3$ slab and a hybrid 53 $\mu$m LiNbO$_3$--49 $\mu$m ErFeO$_3$ are very similar.}
\end{figure}

We recorded the progress of the THz wave in the hybrid slab in a manner similar to that used earlier~\cite{Yang2010}, by exploiting the electro-optic effect wherein the THz electric field $E_{THz}$ modulates the LiNbO$_3$ refractive index~\cite{Winnewisser1997a}. We recorded the time-dependent THz electric-field profile $E_{THz}(t)$ at various positions in the propagation direction ($y$-direction as defined in Fig. \ref{fig:setup}a) by passing the 800-nm probe pulse through the structure and measuring the variably delayed THz-induced depolarization as indicated in Fig.~\ref{fig:setup}a. A display of the profiles $E_{THz}(t)$ measured in successive sample locations appears as a 2D space-time matrix $E_{THz}(y, t)$ as shown in Fig.~\ref{fig:setup}e. Although the 400 nm pump pulse launches counterpropagating waves, initially we only observe the rightward travelling wave because the pump pulse is to the left of the detection window. The first several oscillation cycles of this wave are due to the lowest-order waveguide mode, and the second oscillation cycles are due to the next waveguide mode. Sometime later, we also observe an additional rightward propagating wave starting around 40 ps and leftward propagating wave starting at 90 ps that are due to partial reflections at the edges of the sample. Finally, a 2D Fourier transform was performed to obtain the wavevector-frequency dispersion curve $E_{THz}(k, \omega)$. In Fig.~\ref{fig:dispersion}a, we show the experimentally recorded dispersion curve for the 53 $\mu$m LiNbO$_3$ slab. Owing to the almost linear dispersion of the THz phonon-polaritons in our THz pulse bandwidth, two well separated transverse-electric (TE) dielectric waveguide modes~\cite{Yang2010} appear that are reproduced by a numerical calculation assuming a core refractive index of $n = 5.0$~\cite{Palfalvi2005b} (using the freely available software package MPB~\cite{Johnson2001}). In Fig.~\ref{fig:dispersion}b, we show the experimentally recorded dispersion curves for the hybrid 53 $\mu$m LiNbO$_3$ -- 8 $\mu$m air -- 40 $\mu$m ErFeO$_3$ slab where three well separated modes appear, and largely match numerically calculated solutions that assume an ErFeO$_3$ index of $n= 4.9$~\cite{Kozlov1993}. The existence of three modes in Fig.~\ref{fig:dispersion}b in comparison to two modes in Fig.~\ref{fig:dispersion}a is a result of the increased thickness of the hybrid slab. As demonstrated in Fig. 2c, the waveguide modes become lowered in frequency and more closely spaced as the thickness of the waveguide is increased. We make two simplifications in the proceeding analysis. Firstly, the existence of an air-gap can be ignored as it does not significantly influence the dispersion curves, as shown in Fig. \ref{fig:dispersion}d. Secondly, as demonstrated by Fig. \ref{fig:dispersion}e, at most frequencies the hybrid waveguide dispersion curves resemble those in a 93 $\mu$m slab of LiNbO$_3$. This is because of the very similar refractive indices of LiNbO$_3$ ($n=5.0$) and ErFeO$_3$ ($n=4.9$). With these assumptions, we can focus on the notable feature in Fig. \ref{fig:dispersion}b near the magnon resonance at 0.67 THz: an avoided crossing in both the first and second-order waveguide modes.

\section{Avoided crossing, normal-mode beating, and the band gap}\label{sec:Avoided} 

The avoided crossing is a hallmark signature of strong coupling and can be explained by the formation of new mixed modes, the magnon-phonon-polaritons. As discussed above, the new modes emerge from the strong coupling of the magnon and phonon-polariton modes. A phenomenological approach of coupled oscillators can be used to describe this interaction, using the Hamiltonian given as 

\begin{eqnarray}
H & = & H_{\text{pp}} + H_\text{m} + W_{\text{int}}\label{eq:1},\\
H & = & \hbar \begin{pmatrix} 
\beta c - i \kappa & 0 \\ 0 & 0  
\end{pmatrix} +
\hbar \begin{pmatrix}
0 & 0 \\ 0 & \omega_o - i\gamma
\end{pmatrix}+\nonumber\\
&& \hbar \begin{pmatrix}
0 & \Omega/2 \\ \Omega/2 & 0
\end{pmatrix} \label{eq:2},\\
\omega_{\text{UP/LP}} & = & \frac{\beta c + \omega_o}{2} - i \frac{\gamma + \kappa}{2} \pm \nonumber\\
&& \sqrt{\Omega^2 + \left((\beta c - \omega_o)+i(\gamma-\kappa)\right)^2}\label{eq:3}.
\end{eqnarray}

Here, $H_{\text{pp}}$ \& $H_\text{m}$ are the bare Hamiltonians and $\kappa = 12 $ GHz \& $\gamma = 8$ GHz~\cite{Kozlov1993} are the FWHM linewidths of the phonon-polariton and magnon, respectively. Correspondingly, $W_{int}$ denotes their interaction, $c$ is the speed of light in vacuum, $\omega$ is the eigenfrequency of the coupled system, $\omega_o/2\pi=0.67$ THz is the magnon resonance frequency, $\beta$ is the hybrid waveguide wavevector that acts as the detuning parameter between the two bare modes via the term $\beta c - \omega_o$, and $\Omega$ is the splitting frequency which is a measure of the coupling strength. Note that $\beta$ can be defined in terms of the relation $\beta=n_{\text{eff}}\omega/c$ where $n_{\text{eff}}$ is termed the effective index and is defined as the ratio of the speed of light to the phase velocity of the waveguide mode. $\beta$ thereby contains a frequency and mode dependent weighted average of the refractive indices of the LiNbO$_3$, ErFeO$_3$, and air. The solutions to this $2\times 2$ Hamiltonian correspond to the upper and lower magnon-phonon-polariton branches $\omega_{\text{UP}}$ and $\omega_{\text{LP}}$, respectively.  

\begin{figure*}
\includegraphics[width=172mm]{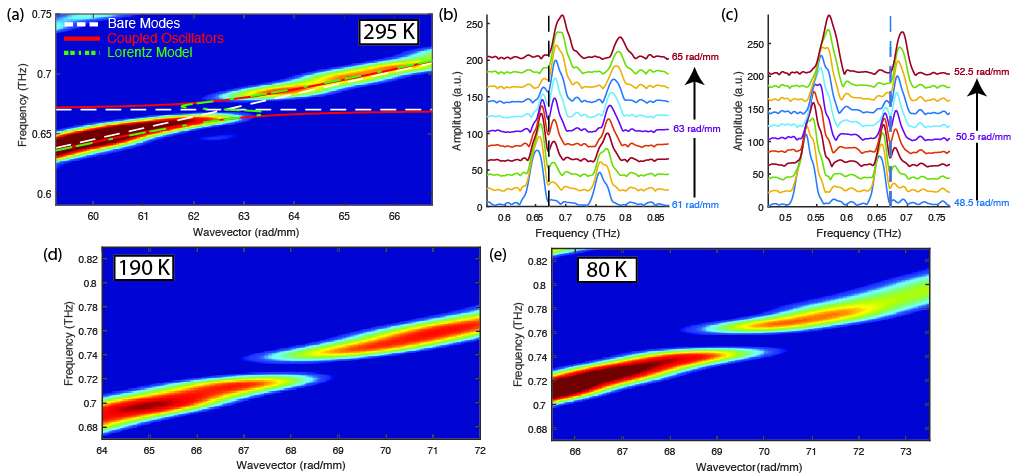}
\caption{\label{fig:avoided} Normal mode splitting of THz magnon-phonon-polaritons. (a) Dispersion curve shows the normal mode splitting between the magnon resonance at 0.67 THz and the first-order phonon-polariton mode. The white dashed curves show the uncoupled magnon frequency and a numerical calculation of the phonon-polariton dispersion, while the red solid and green dash-dotted curves show calculated magnon-phonon-polariton dispersion curves assuming a coupled oscillator model and Lorentz model for the permeability of ErFeO$_3$, respectively. (b) Frequency spectrum at a series of wavevectors spaced by 0.4 rad/mm and ranging from 61 to 65 rad/mm. (c) Frequency spectrum at a series of wavevectors spaced by 0.4 rad/mm and ranging from 48.5 to 52.5 rad/mm. The double-peaked structure in the first and second modes near the black dashed line, which indicates the magnon resonance at 0.67 THz, can be observed in (b) and (c), respectively. (d), (e) Normal mode splitting of THz magnon-phonon-polaritons as a function of temperature. Dispersion curve shows the mode splitting between the first-order phonon-polariton mode and the magnon resonance at (d) 0.73 THz at 190 K and (e) 0.75 THz at 80 K.}
\end{figure*}

In Fig.~\ref{fig:avoided}a, we show an enlargement of the LiNbO$_3$-air-ErFeO$_3$ hybrid slab dispersion curve for the first-order waveguide mode near the magnon resonance, in which the avoided crossing feature can be seen more clearly.  The bare magnon and phonon-polariton modes are only well fitted to the experimental dispersion at large detunings, where the hybrid modes resemble the bare modes (Fig.~\ref{fig:avoided}a, dashed lines). In contrast, the upper and lower magnon-phonon-polariton branches from equation (3) are well fitted at all detunings, reproducing the normal mode splitting (Fig.~\ref{fig:avoided}a, solid lines). As a complement, in Fig.~\ref{fig:avoided}b we show the frequency spectra for a series of wavevectors at various detunings from the magnon resonance. The first-order mode has phonon-polariton character at large detunings but shows a double-peaked structure at and near zero detuning where the magnon and phonon-polariton modes are strongly mixed. Although the second-order mode in Fig.~\ref{fig:avoided}b is far detuned for this range of wavevectors and thus shows a simple monotonic increase in center frequency, it also demonstrates a double-peaked spectrum at zero detuning (see Fig. \ref{eq:3}c). To confirm the avoided crossing was due to the AF-mode magnon in ErFeO$_3$, we also performed temperature dependent experiments as shown in Fig. \ref{fig:avoided}d and \ref{fig:avoided}e. We clearly observe the avoided crossing shift to 0.75 THz at 80 K, in agreement with the shift of the AF-mode magnon frequency in ErFeO$_3$~\cite{Kozlov1993}.

The splitting frequencies for the first and second-order waveguide modes were found to be $\Omega_1/2\pi = 20$ GHz and $\Omega_2/2\pi = 18$ GHz, respectively. Along with the linewidths for the magnon $\gamma/2\pi = 8$ GHz and phonon-polariton mode $\kappa/2\pi = 12$ GHz, these allow us to estimate the cooperativity factor $C = \Omega^2/\kappa\gamma$. The cooperativity factor is a dimensionless quantity that compares the coupling strength to the losses in the system, with $C > 1$ considered strong coupling. The first and second-order phonon-polariton modes couple to the magnons with cooperativity factors $C = 4.0$ and $C = 3.4$, respectively, further validating that the system is in the strong coupling regime.

\section{Lorentz Model}\label{sec:Lorentz}

So far, we have used a classical coupled oscillator model for magnons and phonons because the two can be modelled as damped harmonic oscillators. A Lorentz model (i.e. linear dispersion theory) can also be used to describe the polariton formation. In this case, we assume the permeability of ErFeO$_3$ is modelled as a damped harmonic oscillator near the magnon resonant frequency. As such, the dispersion relation for the electromagnetic wave can be written as 
\begin{eqnarray}
\beta^2c^2 &=& \omega^2\mu_r \label{eq:4}\\
&=& \omega^2\left(1+\frac{\Delta\mu_r\omega_o^2}{\omega_o^2-\omega^2-i\gamma\omega} \right)\label{eq:5}
\end{eqnarray}
where we assume that $\Delta\mu_k=8\times 10^{-4}$~\cite{Kozlov1993}. Using this approach, we obtained a dispersion curve as overlaid on the data in Fig. \ref{fig:avoided}a (dash-dotted lines). Outside the anomalous frequency range, the Lorentz model shows excellent agreement to the data and the coupled oscillator model of Eqns. (\ref{eq:1}-\ref{eq:3}). This is not surprising given that the Lorentz and coupled oscillator models are similar phenomenological approaches to describing strong light-matter coupling. The similarity comes from the fact that the electromagnetic field can be quantized as a harmonic oscillator or defined using dispersion relations. The discrepancy (anomalous dispersion versus an avoided crossing) comes from the fact that the Lorentz model allows for a complex wavevector i.e. spatial damping. This damping is the reason the anomalous region was not seen in our experiments. In fact, in the absence of this damping (i.e. $\gamma\sim0$) and close to resonance (i.e. $\omega\sim\omega_o$), the Lorentz model is equivalent to the coupled oscillator model and can be written as 
\begin{eqnarray}
\omega_{\pm} = \frac{\beta c+\omega_o}{2} + \sqrt{\Delta\mu\omega_o^2+\left(\beta c - \omega_o\right)^2}\label{eq:6}
\end{eqnarray}
This has the same form as Eqn. (\ref{eq:3}) in the absence of damping, and also models an avoided crossing. In this case, the splitting frequency is given by $\sqrt{\Delta\mu_r}\omega_o/2\pi \simeq 19$ GHz, which closely matches the experimentally extracted values for $\Omega_1/2\pi$ and $\Omega_2/2\pi$.  

Despite the convergence of these two models, we should point out that light-matter interactions that allow for light propagation always have an associated complex wavevector, and strong light-matter coupling in these systems always exhibit anomalous dispersion. The Lorentz model is therefore often used to describe polaritons in bulk and waveguide geometries, while the coupled oscillator model is more suitable when the coupling occurs in a photonic cavity for example. Interestingly, while both models are classical, they can also be used to describe the quantum observation of vacuum Rabi splitting~\cite{Reithmaier2004,Zhu1990}.  The ability of these classical models to describe a quantum interaction stems from the fact that for weak atomic excitations, the quantum theory of interaction between an electromagnetic field and an atom is equivalent to the Lorentz model for a classical oscillator. Thus, to earn the distinction of quantum strong coupling, it is not sufficient to observe a splitting. Instead, a dependence of splitting energy on the electric field or number of oscillators must be observed~\cite{Torma2015,Fink2010}. Such a dependence was not observed in our experiments.

We would also like to note that there exist THz-frequency phonon modes in ErFeO$_3$ itself, and it was demonstrated that driving the modes near $\sim 20$ THz can create an effective magnetic field that subsequently drives the AF-mode magnon~\cite{Nova2016}. However, this is a non-resonant Raman-type effect using $\sim$Mv/cm THz field strengths. In contrast, we use the phonons in LiNbO$_3$ to generate THz waves, which then resonantly drive the magnons in ErFeO$_3$ via the magnetic field component of the electromagnetic wave. The driving of the magnons is thus linear in the field. In addition, in our system the magnon response can be resolved through its coupling to the E-field at $\sim$V/cm THz E-field strengths owing to the efficient EO detection in LiNbO$_3$.

\section{Absorption vs. Strong coupling}\label{sec:Absorption}

\begin{figure}
\includegraphics[width=85mm]{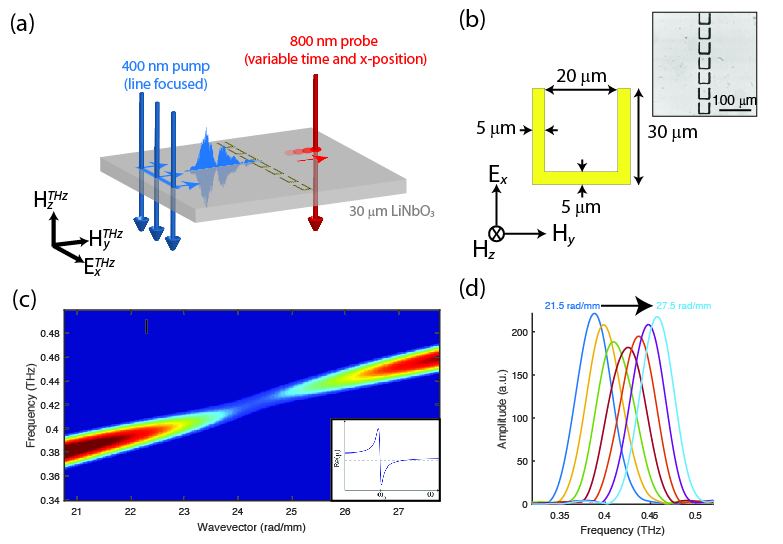}
\caption{\label{fig:absorption} Linear absorption by an array of gold split ring resonators with magnetic resonance frequency of 0.42 THz fabricated on a slab of 30 $\mu$m LiNbO$_3$. (a) Geometry of the experiment demonstrating the acquistion of the spatiotemporal data after transmission through the array of SRRs. Associated polarizations of the THz electromagnetic field are also shown, where the $x$-axis corresponds to the $c$-axis of the LiNbO$_3$ crystal. (b) Dimensions of the SRR and the associated THz E- and H-field polarizations in the experiment. Inset shows a microscope image of the SRR array with periodicity of 40 $\mu$m. (c) Dispersion curve near the magnetic resonance frequency, showing a dip in the absorption on resonance and no evidence of an avoided crossing. Inset shows SRR modelled by a Lorentzian model for the permeability. (d) Frequency spectrum at a series of wavevectors spaced by 1 rad/mm and ranging from 21.5 to 27.5 rad/mm.}
\end{figure}

The features in Fig. \ref{fig:dispersion}b and Fig. \ref{fig:avoided} are not simply “gaps” in the dispersion curve due to linear absorption. To demonstrate this distinction, we have performed an experiment that demonstrates the case of linear absorption due to a magnetic resonance by measuring the dispersion curve after transmission through an array of split-ring resonators (SRRs) patterned onto the surface of a 30 $\mu$m $x$-cut LiNbO$_3$ slab (see Fig. \ref{fig:absorption}a and \ref{fig:absorption}b). The SRRs are 200 nm thick gold on a thin chromium adhesion layer deposited by optical lithography directly on the surface of the LiNbO$_3$ slab. Due to the orientation of the $E$- and $H$- fields in the experiment, the SRRs only exhibit a magnetic resonance that can be modelled using a Lorentz model for the magnetic permeability~\cite{Zhou2007}. Thus, they are an ideal comparison to the AF-mode magnon in ErFeO$_3$.  From the dispersion curve of Fig. \ref{fig:absorption}c, there is no evidence of an avoided crossing. Instead, we only see a reduced amplitude around the resonant frequency. Furthermore, from the wavevector-dependent spectra in Fig. \ref{fig:absorption}d, there is never a double-peaked structure. Instead, there is always a single-peaked structure with the magnitude of the peak reduced when on resonance. This is the behavior expected for linear absorption, and is demonstrably different from the avoided crossing shown in Fig. \ref{fig:avoided}. 

\section{Energy exchange and bandgap}\label{sec:Exchange}

\begin{figure}
\includegraphics[width=85mm]{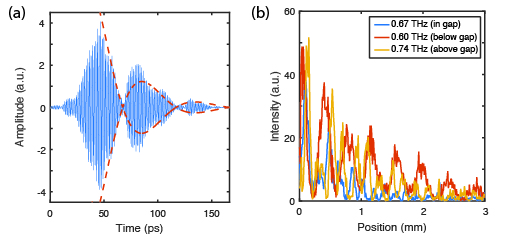}
\caption{\label{fig:dynamics} Temporal and spatial dynamics of THz magnon-phonon-polaritons. (a) Time-domain data obtained by frequency-filtering of the 63 rad/mm spectrum shown in Fig.~\ref{fig:avoided}b to exclude all frequencies outside of the 0.60-0.73 THz  range and inverse Fourier transformation. Normal mode beating between the upper and lower magnon-phonon-polariton branches matches that expected for a phonon-magnon energy exchange period of $2\pi/\Omega=50$ ps as shown by the dashed line. (b) Decay length of 0.48 mm for the mode within the band gap of the avoided crossing is significantly shorter than for modes above or below the band gap, which have decay lengths of 1.27 mm and 1.32 mm, respectively.}
\end{figure}

The strong coupling suggests there is coherent energy exchange between the magnons and phonon-polaritons at a rate $\Omega$. Our spatially and temporally resolved measurements permit direct observation of this process. Figure \ref{fig:dynamics}a shows time-dependent data at a selected wavevector $\beta = 63$ rad/mm, obtained by frequency filtering the spectrum shown in Fig.~\ref{fig:avoided}b around the lower-order waveguide mode and inverse Fourier transforming to return to the time domain. This allows us to examine the evolution of only the lower-order waveguide mode at frequencies around the avoided crossing in order to illustrate the energy exchange clearly. The signal shows beating with energy exchange period $2\pi/\Omega =50$ ps and decay time $\tau  = 2/(\kappa+\gamma) = 100$ ps (see Fig.~\ref{fig:dynamics}a, dashed line).

Although the polariton band gap in the hybrid waveguide excludes any eigenmodes around the bare magnon frequency, the phonon-polariton wavepacket entering from the LiNbO$_3$ extension includes frequencies within this forbidden gap. These waves are evanescent and so decay exponentially into the hybrid slab. Ignoring other sources of absorption, the decay rate should be greatest at the center of the gap (i.e. the bare magnon frequency) and the rate should decrease as frequencies at the gap edges are approached~\cite{Joannopoulos2008}.  As shown in Fig.~\ref{fig:dynamics}b, the frequency component of 0.67 THz in the center of the band gap has a decay length of 0.48 mm, 2-3 times shorter than those either above or below the band gap. The frequency-dependent damping rates outside the forbidden range closely follow those of bare LiNbO$_3$~\footnote{Note that the space-time plot in Fig.~\ref{fig:setup}b, which was used for the eigenmode analysis starts, at approximately 0.5 mm into the hybrid structure in order to avoid significant evanescent wave contributions~\cite{Feurer2007ARM}}. 

\section{Conclusions}\label{sec:Conclusion}

We have shown strong coupling between phonon-polaritons and magnons in a LiNbO$_3$-air-ErFeO$_3$ hybrid slab in the THz frequency range.  In the wavevector-frequency domain, we have observed an avoided crossing in the dispersion curve for both the first and second-order TE dielectric waveguide modes, which exhibit a normal mode splitting of $\Omega_1/2\pi =20$ GHz and $\Omega_2/2\pi = 18$ GHz with corresponding cooperativity factors $C=4.0$ and $C=3.4$, respectively.  Correspondingly, in the time domain we observed a beating between the upper and lower magnon-phonon-polaritons with an oscillation period of $2\pi/\Omega=50$ ps, which is the period for energy exchange between phonons and magnons. In the spatial domain, we observed evanescent wave decay at frequencies within the band gap as the wavepacket entered the hybrid waveguide. These features provide direct evidence for strong coupling of the magnetization and lattice vibrations to THz frequency light, and illustrate a system with multifunctional capabilities. In addition to the various form of electrical control afforded by LiNbO$_3$~\cite{Cho1987}, its large electro-optic constants enables efficient generation and detection of THz waves~\cite{Yeh2007a,Winnewisser1997a} that directly couple to the ErFeO$_3$ magnons, with handles for both spatial and temporal shaping of the THz field profile~\cite{Feurer2007ARM,Feurer2003c}. The coupling to THz light in a waveguide also allows coherent propagation of the magnetization in a highly confined and efficient manner. Ultimately, the on-chip geometry and ultrafast time scales used here may show promise for an avenue toward ultrafast THz spintronics. 

\begin{acknowledgments}
The optical measurements planned and conducted by P.S., J.L., and K.A.N. at MIT were supported as part of the Center for Excitonics, an Energy Frontier Research Center (EFRC), funded by the U.S. Department of Energy (DOE), Office of Science, Basic Energy Sciences (BES), under award number DE-SC0001088. The crystal growth and preparation by M.X., W.R., and S.C. at Shanghai University and S.K. at the Academy of Sciences of the Czech Republic were respectively supported by the National Natural Science Foundation of China (NSFC) grant No. 51372149 and No. 51672171 and  the Czech Science Foundation Project No. 15-08389S. P. S was supported in part by a Postgraduate Scholarship (PGS-D) from the Natural Sciences and Engineering Research Council of Canada (NSERC).
\end{acknowledgments}

\appendix

\section{Energy distribution between lattice and electromagnetic wave}\label{sec:latticeenergy}

To calculate the energy distribution between the lattice motion and electromagnetic field, we begin by describing the coupling between the transverse polar optic phonon mode and the electromagnetic field. The coupled equations are given as~\cite{Huang1951a} 

\begin{eqnarray}
P &=& \omega_{TO}\sqrt{\epsilon_o(\varepsilon_o-\varepsilon_\infty)}Q+\epsilon_o(\varepsilon_\infty-1)E, \label{eq:S1}\\
\ddot{Q} &=& -\omega_{TO}^2Q-\Gamma\dot{Q} + \omega_{TO}\sqrt{\epsilon_o(\varepsilon_o-\varepsilon_\infty)}E.\label{eq:S2}
\end{eqnarray}

Here, $P$ is the macroscopic polarization of the material, and $Q$ is the normalized ionic displacement
of the polar optic phonon mode where an over-dot indicates the time derivative. In addition, $\omega_{TO}$ is the transverse optic phonon frequency, $\epsilon_o$ is vacuum permittivity, and $\varepsilon_\infty$ and $\varepsilon_o$ are the high- and low-frequency dielectric constants of the material, respectively. We now introduce Poynting's theorem, which relates the energy stored in the electromagnetic field to the work done on the electric dipoles in the medium. In differential form, the energy balance can be expressed as
\begin{equation}
-\nabla\cdot S = \dot{W}.\label{eq:S3}
\end{equation}

where $S$ is the Poynting vector and $W$ is the energy density.  Equation (\ref{eq:S3}) states that the rate of energy decrease in the medium (RHS) is equal to the amount of energy flow out of the medium (LHS). To determine the energy distribution between the lattice and electromagnetic field, we must determine an appropriate energy density that satisfies Eqn. (\ref{eq:S3}). Such an energy density was proposed by Huang~\cite{Huang1951a} in the absence of damping ($\Gamma = 0 $), which is given as
\begin{equation}
W = \frac{1}{2}\left(\dot{Q}^2+\omega_{TO}^2Q^2\right)+\frac{1}{2}\left( \epsilon_o \varepsilon_\infty E^2 + \mu_o H^2\right).\label{eq:S4}
\end{equation}

Where $E$ and $H$ are the electric and magnetic fields, respectively, and $Q$ is the normalized displacement of the phonon mode, and $\mu_o$ is the vacuum permeability. The first term in brackets of Eqn. (\ref{eq:S4}) represents the vibrational energy of the TO phonon mode, given as a sum of the potential and kinetic energy terms one normally uses for a harmonic oscillator. The second term in brackets represents the electromagnetic energy, with an inclusion of the electronic response via the term $\varepsilon_\infty$. As such, we can write the time-averaged energies in the lattice and electromagnetic wave as 
\begin{eqnarray}
\langle W_{latt} \rangle & = & \frac{\omega_{TO}^2\epsilon_o(\varepsilon_o-\varepsilon_\infty)(\omega^2+\omega_{TO}^2)}{4(\omega_{TO}^2-\omega^2)^2}\left|E_o\right|^2,\label{eq:S5}\\
\langle W_{EM} \rangle & = &\frac{1}{4}\left(\epsilon_o\varepsilon_\infty+\frac{1}{\mu_o}\left(\frac{k}{\omega}\right)^2\right)\left|E_o\right|^2.\label{eq:S6}
\end{eqnarray}

The time-averaged fraction of mechanical energy in the mode is then given as 
\begin{eqnarray}
\langle \rho \rangle = \frac{\langle W_{latt} \rangle}{\langle W_{latt} \rangle+\langle W_{EM} \rangle}
\end{eqnarray}

Where $\langle W_{latt} \rangle$ and $\langle W_{EM} \rangle$ are the time averaged lattice and electromagnetic energies, respectively. Using Eqns. (\ref{eq:S5}) and (\ref{eq:S6}) and assuming material parameters for LN as specified in the work of Feurer et al.~\cite{Feurer2007ARM} , the fraction of energy in the lattice is 32\% at 1 THz.
\nocite{*}
%


\begin{thebibliography}{93}%
\makeatletter
\providecommand \@ifxundefined [1]{%
 \@ifx{#1\undefined}
}%
\providecommand \@ifnum [1]{%
 \ifnum #1\expandafter \@firstoftwo
 \else \expandafter \@secondoftwo
 \fi
}%
\providecommand \@ifx [1]{%
 \ifx #1\expandafter \@firstoftwo
 \else \expandafter \@secondoftwo
 \fi
}%
\providecommand \natexlab [1]{#1}%
\providecommand \enquote  [1]{``#1''}%
\providecommand \bibnamefont  [1]{#1}%
\providecommand \bibfnamefont [1]{#1}%
\providecommand \citenamefont [1]{#1}%
\providecommand \href@noop [0]{\@secondoftwo}%
\providecommand \href [0]{\begingroup \@sanitize@url \@href}%
\providecommand \@href[1]{\@@startlink{#1}\@@href}%
\providecommand \@@href[1]{\endgroup#1\@@endlink}%
\providecommand \@sanitize@url [0]{\catcode `\\12\catcode `\$12\catcode
  `\&12\catcode `\#12\catcode `\^12\catcode `\_12\catcode `\%12\relax}%
\providecommand \@@startlink[1]{}%
\providecommand \@@endlink[0]{}%
\providecommand \url  [0]{\begingroup\@sanitize@url \@url }%
\providecommand \@url [1]{\endgroup\@href {#1}{\urlprefix }}%
\providecommand \urlprefix  [0]{URL }%
\providecommand \Eprint [0]{\href }%
\providecommand \doibase [0]{http://dx.doi.org/}%
\providecommand \selectlanguage [0]{\@gobble}%
\providecommand \bibinfo  [0]{\@secondoftwo}%
\providecommand \bibfield  [0]{\@secondoftwo}%
\providecommand \translation [1]{[#1]}%
\providecommand \BibitemOpen [0]{}%
\providecommand \bibitemStop [0]{}%
\providecommand \bibitemNoStop [0]{.\EOS\space}%
\providecommand \EOS [0]{\spacefactor3000\relax}%
\providecommand \BibitemShut  [1]{\csname bibitem#1\endcsname}%
\let\auto@bib@innerbib\@empty
\bibitem [{\citenamefont {Kwek}\ \emph {et~al.}(2014)\citenamefont {Kwek},
  \citenamefont {Auff{\`{e}}ves}, \citenamefont {Gerace}, \citenamefont
  {Richard}, \citenamefont {Portolan}, \citenamefont {Santos}, \citenamefont
  {Kwek},\ and\ \citenamefont {Miniatura}}]{Kwek2014}%
  \BibitemOpen
  \bibfield  {author} {\bibinfo {author} {\bibfnamefont {L.~C.}\ \bibnamefont
  {Kwek}}, \bibinfo {author} {\bibfnamefont {A.}~\bibnamefont
  {Auff{\`{e}}ves}}, \bibinfo {author} {\bibfnamefont {D.}~\bibnamefont
  {Gerace}}, \bibinfo {author} {\bibfnamefont {M.}~\bibnamefont {Richard}},
  \bibinfo {author} {\bibfnamefont {S.}~\bibnamefont {Portolan}}, \bibinfo
  {author} {\bibfnamefont {M.~F.}\ \bibnamefont {Santos}}, \bibinfo {author}
  {\bibfnamefont {L.~C.}\ \bibnamefont {Kwek}}, \ and\ \bibinfo {author}
  {\bibfnamefont {C.}~\bibnamefont {Miniatura}},\ }\href@noop {} {\emph
  {\bibinfo {title} {{Strong Light-matter Coupling: From Atoms to Solid-state
  Systems}}}}\ (\bibinfo  {publisher} {World Scientific, New Jersey},\ \bibinfo
  {address} {New Jersey},\ \bibinfo {year} {2014})\BibitemShut {NoStop}%
\bibitem [{\citenamefont {Mills}\ and\ \citenamefont
  {Burstein}(2001)}]{Mills2001}%
  \BibitemOpen
  \bibfield  {author} {\bibinfo {author} {\bibfnamefont {D.~L.}\ \bibnamefont
  {Mills}}\ and\ \bibinfo {author} {\bibfnamefont {E.}~\bibnamefont
  {Burstein}},\ }\href@noop {} {\bibfield  {journal} {\bibinfo  {journal}
  {Reports on Progress in Physics}\ }\textbf {\bibinfo {volume} {37}},\
  \bibinfo {pages} {817} (\bibinfo {year} {2001})}\BibitemShut {NoStop}%
\bibitem [{\citenamefont {Snoke}(2015)}]{Snoke2015}%
  \BibitemOpen
  \bibfield  {author} {\bibinfo {author} {\bibfnamefont {D.}~\bibnamefont
  {Snoke}},\ }\href {https://arxiv.org/abs/1509.01468} {\ ,\ \bibinfo {pages}
  {Preprint at https://arxiv.org/abs/1509.01468} (\bibinfo {year}
  {2015})}\BibitemShut {NoStop}%
\bibitem [{\citenamefont {T{\"{o}}rm{\"{a}}}\ and\ \citenamefont
  {Barnes}(2015)}]{Torma2015}%
  \BibitemOpen
  \bibfield  {author} {\bibinfo {author} {\bibfnamefont {P.}~\bibnamefont
  {T{\"{o}}rm{\"{a}}}}\ and\ \bibinfo {author} {\bibfnamefont {W.~L.}\
  \bibnamefont {Barnes}},\ }\href@noop {} {\bibfield  {journal} {\bibinfo
  {journal} {Reports on progress in physics. Physical Society (Great Britain)}\
  }\textbf {\bibinfo {volume} {78}},\ \bibinfo {pages} {013901} (\bibinfo
  {year} {2015})}\BibitemShut {NoStop}%
\bibitem [{\citenamefont {Aspelmeyer}\ \emph {et~al.}(2014)\citenamefont
  {Aspelmeyer}, \citenamefont {Kippenberg},\ and\ \citenamefont
  {Marquardt}}]{Aspelmeyer2014}%
  \BibitemOpen
  \bibfield  {author} {\bibinfo {author} {\bibfnamefont {M.}~\bibnamefont
  {Aspelmeyer}}, \bibinfo {author} {\bibfnamefont {T.~J.}\ \bibnamefont
  {Kippenberg}}, \ and\ \bibinfo {author} {\bibfnamefont {F.}~\bibnamefont
  {Marquardt}},\ }\href {\doibase 10.1103/RevModPhys.86.1391} {\bibfield
  {journal} {\bibinfo  {journal} {Reviews of Modern Physics}\ }\textbf
  {\bibinfo {volume} {86}},\ \bibinfo {pages} {1391} (\bibinfo {year}
  {2014})},\ \Eprint {http://arxiv.org/abs/0712.1618} {arXiv:0712.1618}
  \BibitemShut {NoStop}%
\bibitem [{\citenamefont {Barnes}\ \emph {et~al.}(2003)\citenamefont {Barnes},
  \citenamefont {Dereux},\ and\ \citenamefont {Ebbesen}}]{Barnes2003}%
  \BibitemOpen
  \bibfield  {author} {\bibinfo {author} {\bibfnamefont {W.~L.}\ \bibnamefont
  {Barnes}}, \bibinfo {author} {\bibfnamefont {A.}~\bibnamefont {Dereux}}, \
  and\ \bibinfo {author} {\bibfnamefont {T.~W.}\ \bibnamefont {Ebbesen}},\
  }\href {\doibase 10.1038/nature01937} {\bibfield  {journal} {\bibinfo
  {journal} {Nature}\ }\textbf {\bibinfo {volume} {424}},\ \bibinfo {pages}
  {824} (\bibinfo {year} {2003})},\ \Eprint {http://arxiv.org/abs/1312.6806}
  {arXiv:1312.6806} \BibitemShut {NoStop}%
\bibitem [{\citenamefont {Imamoglu}\ \emph {et~al.}(1999)\citenamefont
  {Imamoglu}, \citenamefont {Awschalom}, \citenamefont {Burkard}, \citenamefont
  {DiVincenzo}, \citenamefont {Loss}, \citenamefont {Sherwin},\ and\
  \citenamefont {Small}}]{Imamoglu1999}%
  \BibitemOpen
  \bibfield  {author} {\bibinfo {author} {\bibfnamefont {a.}~\bibnamefont
  {Imamoglu}}, \bibinfo {author} {\bibfnamefont {D.}~\bibnamefont {Awschalom}},
  \bibinfo {author} {\bibfnamefont {G.}~\bibnamefont {Burkard}}, \bibinfo
  {author} {\bibfnamefont {D.~P.}\ \bibnamefont {DiVincenzo}}, \bibinfo
  {author} {\bibfnamefont {D.}~\bibnamefont {Loss}}, \bibinfo {author}
  {\bibfnamefont {M.}~\bibnamefont {Sherwin}}, \ and\ \bibinfo {author}
  {\bibfnamefont {A.}~\bibnamefont {Small}},\ }\href@noop {} {\bibfield
  {journal} {\bibinfo  {journal} {Physical Review Letters}\ }\textbf {\bibinfo
  {volume} {83}},\ \bibinfo {pages} {4204} (\bibinfo {year}
  {1999})}\BibitemShut {NoStop}%
\bibitem [{\citenamefont {Monroe}(2002)}]{Monroe2002}%
  \BibitemOpen
  \bibfield  {author} {\bibinfo {author} {\bibfnamefont {C.}~\bibnamefont
  {Monroe}},\ }\href {\doibase 10.1038/416238a} {\bibfield  {journal} {\bibinfo
   {journal} {Nature}\ }\textbf {\bibinfo {volume} {416}},\ \bibinfo {pages}
  {238} (\bibinfo {year} {2002})},\ \Eprint
  {http://arxiv.org/abs/arXiv:1011.4115} {arXiv:arXiv:1011.4115} \BibitemShut
  {NoStop}%
\bibitem [{\citenamefont {You}\ and\ \citenamefont {Nori}(2003)}]{You2003}%
  \BibitemOpen
  \bibfield  {author} {\bibinfo {author} {\bibfnamefont {J.~Q.}\ \bibnamefont
  {You}}\ and\ \bibinfo {author} {\bibfnamefont {F.}~\bibnamefont {Nori}},\
  }\href {\doibase 10.1103/PhysRevB.68.064509} {\bibfield  {journal} {\bibinfo
  {journal} {Physical Review B}\ }\textbf {\bibinfo {volume} {68}},\ \bibinfo
  {pages} {064509} (\bibinfo {year} {2003})},\ \Eprint
  {http://arxiv.org/abs/0306207} {arXiv:0306207 [cond-mat]} \BibitemShut
  {NoStop}%
\bibitem [{\citenamefont {Thompson}\ \emph {et~al.}(1992)\citenamefont
  {Thompson}, \citenamefont {Rempe},\ and\ \citenamefont
  {Kimble}}]{Thompson1992}%
  \BibitemOpen
  \bibfield  {author} {\bibinfo {author} {\bibfnamefont {R.~J.}\ \bibnamefont
  {Thompson}}, \bibinfo {author} {\bibfnamefont {G.}~\bibnamefont {Rempe}}, \
  and\ \bibinfo {author} {\bibfnamefont {H.~J.}\ \bibnamefont {Kimble}},\
  }\href {\doibase 10.1103/PhysRevLett.68.1132} {\bibfield  {journal} {\bibinfo
   {journal} {Physical Review Letters}\ }\textbf {\bibinfo {volume} {68}},\
  \bibinfo {pages} {1132} (\bibinfo {year} {1992})}\BibitemShut {NoStop}%
\bibitem [{\citenamefont {Wallraff}\ \emph
  {et~al.}(2004{\natexlab{a}})\citenamefont {Wallraff}, \citenamefont
  {Schuster}, \citenamefont {Blais}, \citenamefont {Frunzio}, \citenamefont
  {Huang}, \citenamefont {Majer}, \citenamefont {Kumar}, \citenamefont
  {Girvin},\ and\ \citenamefont {Schoelkopf}}]{Wallraff2004}%
  \BibitemOpen
  \bibfield  {author} {\bibinfo {author} {\bibfnamefont {A.}~\bibnamefont
  {Wallraff}}, \bibinfo {author} {\bibfnamefont {D.}~\bibnamefont {Schuster}},
  \bibinfo {author} {\bibfnamefont {A.}~\bibnamefont {Blais}}, \bibinfo
  {author} {\bibfnamefont {L.}~\bibnamefont {Frunzio}}, \bibinfo {author}
  {\bibfnamefont {R.}~\bibnamefont {Huang}}, \bibinfo {author} {\bibfnamefont
  {J.}~\bibnamefont {Majer}}, \bibinfo {author} {\bibfnamefont
  {S.}~\bibnamefont {Kumar}}, \bibinfo {author} {\bibfnamefont
  {S.}~\bibnamefont {Girvin}}, \ and\ \bibinfo {author} {\bibfnamefont
  {R.}~\bibnamefont {Schoelkopf}},\ }\href@noop {} {\bibfield  {journal}
  {\bibinfo  {journal} {Nature}\ }\textbf {\bibinfo {volume} {431}},\ \bibinfo
  {pages} {162} (\bibinfo {year} {2004}{\natexlab{a}})}\BibitemShut {NoStop}%
\bibitem [{\citenamefont {Bellessa}\ \emph {et~al.}(2004)\citenamefont
  {Bellessa}, \citenamefont {Bonnand}, \citenamefont {Plenet},\ and\
  \citenamefont {Mugnier}}]{Bellessa2004}%
  \BibitemOpen
  \bibfield  {author} {\bibinfo {author} {\bibfnamefont {J.}~\bibnamefont
  {Bellessa}}, \bibinfo {author} {\bibfnamefont {C.}~\bibnamefont {Bonnand}},
  \bibinfo {author} {\bibfnamefont {J.~C.}\ \bibnamefont {Plenet}}, \ and\
  \bibinfo {author} {\bibfnamefont {J.}~\bibnamefont {Mugnier}},\ }\href@noop
  {} {\bibfield  {journal} {\bibinfo  {journal} {Physical Review Letters}\
  }\textbf {\bibinfo {volume} {93}},\ \bibinfo {pages} {036404} (\bibinfo
  {year} {2004})}\BibitemShut {NoStop}%
\bibitem [{\citenamefont {Kasprzak}\ \emph {et~al.}(2006)\citenamefont
  {Kasprzak}, \citenamefont {Richard}, \citenamefont {Kundermann},
  \citenamefont {Baas}, \citenamefont {Jeambrun}, \citenamefont {Keeling},
  \citenamefont {Marchetti}, \citenamefont {Szyma{\'{n}}ska}, \citenamefont
  {Andr{\'{e}}}, \citenamefont {Staehli}, \citenamefont {Savona}, \citenamefont
  {Littlewood}, \citenamefont {Deveaud},\ and\ \citenamefont
  {Dang}}]{Kasprzak2006a}%
  \BibitemOpen
  \bibfield  {author} {\bibinfo {author} {\bibfnamefont {J.}~\bibnamefont
  {Kasprzak}}, \bibinfo {author} {\bibfnamefont {M.}~\bibnamefont {Richard}},
  \bibinfo {author} {\bibfnamefont {S.}~\bibnamefont {Kundermann}}, \bibinfo
  {author} {\bibfnamefont {a.}~\bibnamefont {Baas}}, \bibinfo {author}
  {\bibfnamefont {P.}~\bibnamefont {Jeambrun}}, \bibinfo {author}
  {\bibfnamefont {J.~M.~J.}\ \bibnamefont {Keeling}}, \bibinfo {author}
  {\bibfnamefont {F.~M.}\ \bibnamefont {Marchetti}}, \bibinfo {author}
  {\bibfnamefont {M.~H.}\ \bibnamefont {Szyma{\'{n}}ska}}, \bibinfo {author}
  {\bibfnamefont {R.}~\bibnamefont {Andr{\'{e}}}}, \bibinfo {author}
  {\bibfnamefont {J.~L.}\ \bibnamefont {Staehli}}, \bibinfo {author}
  {\bibfnamefont {V.}~\bibnamefont {Savona}}, \bibinfo {author} {\bibfnamefont
  {P.~B.}\ \bibnamefont {Littlewood}}, \bibinfo {author} {\bibfnamefont
  {B.}~\bibnamefont {Deveaud}}, \ and\ \bibinfo {author} {\bibfnamefont
  {L.~S.}\ \bibnamefont {Dang}},\ }\href {\doibase 10.1038/nature05131}
  {\bibfield  {journal} {\bibinfo  {journal} {Nature}\ }\textbf {\bibinfo
  {volume} {443}},\ \bibinfo {pages} {409} (\bibinfo {year}
  {2006})}\BibitemShut {NoStop}%
\bibitem [{\citenamefont {Reithmaier}\ \emph {et~al.}(2004)\citenamefont
  {Reithmaier}, \citenamefont {Sek}, \citenamefont {L{\"{o}}ffler},
  \citenamefont {Hofmann}, \citenamefont {Kuhn}, \citenamefont {Reitzenstein},
  \citenamefont {Keldysh}, \citenamefont {Kulakovskii}, \citenamefont
  {Reinecke},\ and\ \citenamefont {Forchel}}]{Reithmaier2004}%
  \BibitemOpen
  \bibfield  {author} {\bibinfo {author} {\bibfnamefont {J.~P.}\ \bibnamefont
  {Reithmaier}}, \bibinfo {author} {\bibfnamefont {G.}~\bibnamefont {Sek}},
  \bibinfo {author} {\bibfnamefont {A.}~\bibnamefont {L{\"{o}}ffler}}, \bibinfo
  {author} {\bibfnamefont {C.}~\bibnamefont {Hofmann}}, \bibinfo {author}
  {\bibfnamefont {S.}~\bibnamefont {Kuhn}}, \bibinfo {author} {\bibfnamefont
  {S.}~\bibnamefont {Reitzenstein}}, \bibinfo {author} {\bibfnamefont {L.~V.}\
  \bibnamefont {Keldysh}}, \bibinfo {author} {\bibfnamefont {V.~D.}\
  \bibnamefont {Kulakovskii}}, \bibinfo {author} {\bibfnamefont {T.~L.}\
  \bibnamefont {Reinecke}}, \ and\ \bibinfo {author} {\bibfnamefont
  {A.}~\bibnamefont {Forchel}},\ }\href {\doibase 10.1038/nature02969}
  {\bibfield  {journal} {\bibinfo  {journal} {Nature}\ }\textbf {\bibinfo
  {volume} {432}},\ \bibinfo {pages} {197} (\bibinfo {year} {2004})},\ \Eprint
  {http://arxiv.org/abs/nature02969} {arXiv:nature02969 [10.1038]} \BibitemShut
  {NoStop}%
\bibitem [{\citenamefont {Yoshie}\ \emph {et~al.}(2004)\citenamefont {Yoshie},
  \citenamefont {Scherer}, \citenamefont {Hendrickson}, \citenamefont
  {Khitrova}, \citenamefont {Gibbs}, \citenamefont {Rupper}, \citenamefont
  {Ell}, \citenamefont {Shchekin},\ and\ \citenamefont {Deppe}}]{Yoshie2004}%
  \BibitemOpen
  \bibfield  {author} {\bibinfo {author} {\bibfnamefont {T.}~\bibnamefont
  {Yoshie}}, \bibinfo {author} {\bibfnamefont {A.}~\bibnamefont {Scherer}},
  \bibinfo {author} {\bibfnamefont {J.}~\bibnamefont {Hendrickson}}, \bibinfo
  {author} {\bibfnamefont {G.}~\bibnamefont {Khitrova}}, \bibinfo {author}
  {\bibfnamefont {H.~M.}\ \bibnamefont {Gibbs}}, \bibinfo {author}
  {\bibfnamefont {G.}~\bibnamefont {Rupper}}, \bibinfo {author} {\bibfnamefont
  {C.}~\bibnamefont {Ell}}, \bibinfo {author} {\bibfnamefont {O.~B.}\
  \bibnamefont {Shchekin}}, \ and\ \bibinfo {author} {\bibfnamefont {D.~G.}\
  \bibnamefont {Deppe}},\ }\href {\doibase 10.1038/nature02969.1.} {\bibfield
  {journal} {\bibinfo  {journal} {Nature}\ }\textbf {\bibinfo {volume} {432}},\
  \bibinfo {pages} {200} (\bibinfo {year} {2004})}\BibitemShut {NoStop}%
\bibitem [{\citenamefont {Kubo}\ \emph {et~al.}(2010)\citenamefont {Kubo},
  \citenamefont {Ong}, \citenamefont {Bertet}, \citenamefont {Vion},
  \citenamefont {Jacques}, \citenamefont {Zheng}, \citenamefont {Dr{\'{e}}au},
  \citenamefont {Roch}, \citenamefont {Auffeves}, \citenamefont {Jelezko},
  \citenamefont {Wrachtrup}, \citenamefont {Barthe}, \citenamefont {Bergonzo},\
  and\ \citenamefont {Esteve}}]{Kubo2010}%
  \BibitemOpen
  \bibfield  {author} {\bibinfo {author} {\bibfnamefont {Y.}~\bibnamefont
  {Kubo}}, \bibinfo {author} {\bibfnamefont {F.~R.}\ \bibnamefont {Ong}},
  \bibinfo {author} {\bibfnamefont {P.}~\bibnamefont {Bertet}}, \bibinfo
  {author} {\bibfnamefont {D.}~\bibnamefont {Vion}}, \bibinfo {author}
  {\bibfnamefont {V.}~\bibnamefont {Jacques}}, \bibinfo {author} {\bibfnamefont
  {D.}~\bibnamefont {Zheng}}, \bibinfo {author} {\bibfnamefont
  {A.}~\bibnamefont {Dr{\'{e}}au}}, \bibinfo {author} {\bibfnamefont {J.-F.}\
  \bibnamefont {Roch}}, \bibinfo {author} {\bibfnamefont {A.}~\bibnamefont
  {Auffeves}}, \bibinfo {author} {\bibfnamefont {F.}~\bibnamefont {Jelezko}},
  \bibinfo {author} {\bibfnamefont {J.}~\bibnamefont {Wrachtrup}}, \bibinfo
  {author} {\bibfnamefont {M.~F.}\ \bibnamefont {Barthe}}, \bibinfo {author}
  {\bibfnamefont {P.}~\bibnamefont {Bergonzo}}, \ and\ \bibinfo {author}
  {\bibfnamefont {D.}~\bibnamefont {Esteve}},\ }\href {\doibase
  10.1103/PhysRevLett.105.140502} {\bibfield  {journal} {\bibinfo  {journal}
  {Physical review letters}\ }\textbf {\bibinfo {volume} {105}},\ \bibinfo
  {pages} {140502} (\bibinfo {year} {2010})}\BibitemShut {NoStop}%
\bibitem [{\citenamefont {Tabuchi}\ \emph {et~al.}(2014)\citenamefont
  {Tabuchi}, \citenamefont {Ishino}, \citenamefont {Ishikawa}, \citenamefont
  {Yamazaki}, \citenamefont {Usami},\ and\ \citenamefont
  {Nakamura}}]{Tabuchi2014}%
  \BibitemOpen
  \bibfield  {author} {\bibinfo {author} {\bibfnamefont {Y.}~\bibnamefont
  {Tabuchi}}, \bibinfo {author} {\bibfnamefont {S.}~\bibnamefont {Ishino}},
  \bibinfo {author} {\bibfnamefont {T.}~\bibnamefont {Ishikawa}}, \bibinfo
  {author} {\bibfnamefont {R.}~\bibnamefont {Yamazaki}}, \bibinfo {author}
  {\bibfnamefont {K.}~\bibnamefont {Usami}}, \ and\ \bibinfo {author}
  {\bibfnamefont {Y.}~\bibnamefont {Nakamura}},\ }\href {\doibase
  10.1103/PhysRevLett.113.083603} {\bibfield  {journal} {\bibinfo  {journal}
  {Physical Review Letters}\ }\textbf {\bibinfo {volume} {113}},\ \bibinfo
  {pages} {083603} (\bibinfo {year} {2014})},\ \Eprint
  {http://arxiv.org/abs/1405.1913} {arXiv:1405.1913} \BibitemShut {NoStop}%
\bibitem [{\citenamefont {Zhang}\ \emph {et~al.}(2014)\citenamefont {Zhang},
  \citenamefont {Zou}, \citenamefont {Jiang},\ and\ \citenamefont
  {Tang}}]{Zhang2014}%
  \BibitemOpen
  \bibfield  {author} {\bibinfo {author} {\bibfnamefont {X.}~\bibnamefont
  {Zhang}}, \bibinfo {author} {\bibfnamefont {C.~L.}\ \bibnamefont {Zou}},
  \bibinfo {author} {\bibfnamefont {L.}~\bibnamefont {Jiang}}, \ and\ \bibinfo
  {author} {\bibfnamefont {H.~X.}\ \bibnamefont {Tang}},\ }\href@noop {}
  {\bibfield  {journal} {\bibinfo  {journal} {Physical Review Letters}\
  }\textbf {\bibinfo {volume} {113}},\ \bibinfo {pages} {156401} (\bibinfo
  {year} {2014})}\BibitemShut {NoStop}%
\bibitem [{\citenamefont {Awschalom}\ \emph {et~al.}(2013)\citenamefont
  {Awschalom}, \citenamefont {Bassett}, \citenamefont {Dzurak}, \citenamefont
  {Hu},\ and\ \citenamefont {Petta}}]{Awschalom2013}%
  \BibitemOpen
  \bibfield  {author} {\bibinfo {author} {\bibfnamefont {D.~D.~D.}\
  \bibnamefont {Awschalom}}, \bibinfo {author} {\bibfnamefont {L.~L.~C.}\
  \bibnamefont {Bassett}}, \bibinfo {author} {\bibfnamefont {A.~S.~A.}\
  \bibnamefont {Dzurak}}, \bibinfo {author} {\bibfnamefont {E.~L.}\
  \bibnamefont {Hu}}, \ and\ \bibinfo {author} {\bibfnamefont {J.~R.}\
  \bibnamefont {Petta}},\ }\href {\doibase 10.1126/science.1231364} {\bibfield
  {journal} {\bibinfo  {journal} {Science (New York, N.Y.)}\ }\textbf {\bibinfo
  {volume} {339}},\ \bibinfo {pages} {1174} (\bibinfo {year}
  {2013})}\BibitemShut {NoStop}%
\bibitem [{\citenamefont {Chumak}\ \emph {et~al.}(2012)\citenamefont {Chumak},
  \citenamefont {Serga}, \citenamefont {Jungfleisch}, \citenamefont {Neb},
  \citenamefont {Bozhko}, \citenamefont {Tiberkevich},\ and\ \citenamefont
  {Hillebrands}}]{Chumak2012}%
  \BibitemOpen
  \bibfield  {author} {\bibinfo {author} {\bibfnamefont {A.~V.}\ \bibnamefont
  {Chumak}}, \bibinfo {author} {\bibfnamefont {A.~A.}\ \bibnamefont {Serga}},
  \bibinfo {author} {\bibfnamefont {M.~B.}\ \bibnamefont {Jungfleisch}},
  \bibinfo {author} {\bibfnamefont {R.}~\bibnamefont {Neb}}, \bibinfo {author}
  {\bibfnamefont {D.~A.}\ \bibnamefont {Bozhko}}, \bibinfo {author}
  {\bibfnamefont {V.~S.}\ \bibnamefont {Tiberkevich}}, \ and\ \bibinfo {author}
  {\bibfnamefont {B.}~\bibnamefont {Hillebrands}},\ }\href {\doibase
  10.1063/1.3689787} {\bibfield  {journal} {\bibinfo  {journal} {Applied
  Physics Letters}\ }\textbf {\bibinfo {volume} {100}},\ \bibinfo {pages}
  {082405} (\bibinfo {year} {2012})},\ \Eprint {http://arxiv.org/abs/1112.4969}
  {arXiv:1112.4969} \BibitemShut {NoStop}%
\bibitem [{\citenamefont {Kajiwara}\ \emph {et~al.}(2010)\citenamefont
  {Kajiwara}, \citenamefont {Harii}, \citenamefont {Takahashi}, \citenamefont
  {Ohe}, \citenamefont {Uchida}, \citenamefont {Mizuguchi}, \citenamefont
  {Umezawa}, \citenamefont {Kawai}, \citenamefont {Ando}, \citenamefont
  {Takanashi}, \citenamefont {Maekawa},\ and\ \citenamefont
  {Saitoh}}]{Kajiwara2010}%
  \BibitemOpen
  \bibfield  {author} {\bibinfo {author} {\bibfnamefont {Y.}~\bibnamefont
  {Kajiwara}}, \bibinfo {author} {\bibfnamefont {K.}~\bibnamefont {Harii}},
  \bibinfo {author} {\bibfnamefont {S.}~\bibnamefont {Takahashi}}, \bibinfo
  {author} {\bibfnamefont {J.}~\bibnamefont {Ohe}}, \bibinfo {author}
  {\bibfnamefont {K.}~\bibnamefont {Uchida}}, \bibinfo {author} {\bibfnamefont
  {M.}~\bibnamefont {Mizuguchi}}, \bibinfo {author} {\bibfnamefont
  {H.}~\bibnamefont {Umezawa}}, \bibinfo {author} {\bibfnamefont
  {H.}~\bibnamefont {Kawai}}, \bibinfo {author} {\bibfnamefont
  {K.}~\bibnamefont {Ando}}, \bibinfo {author} {\bibfnamefont {K.}~\bibnamefont
  {Takanashi}}, \bibinfo {author} {\bibfnamefont {S.}~\bibnamefont {Maekawa}},
  \ and\ \bibinfo {author} {\bibfnamefont {E.}~\bibnamefont {Saitoh}},\ }\href
  {\doibase 10.1038/nature08876} {\bibfield  {journal} {\bibinfo  {journal}
  {Nature}\ }\textbf {\bibinfo {volume} {464}},\ \bibinfo {pages} {262}
  (\bibinfo {year} {2010})},\ \Eprint {http://arxiv.org/abs/1302.1352}
  {arXiv:1302.1352} \BibitemShut {NoStop}%
\bibitem [{\citenamefont {Ivanov}(2014)}]{Ivanov2014}%
  \BibitemOpen
  \bibfield  {author} {\bibinfo {author} {\bibfnamefont {B.~A.}\ \bibnamefont
  {Ivanov}},\ }\href@noop {} {\enquote {\bibinfo {title} {{Spin dynamics of
  antiferromagnets under action of femtosecond laser pulses (Review
  Article)}},}\ } (\bibinfo {year} {2014})\BibitemShut {NoStop}%
\bibitem [{\citenamefont {Talbayev}\ \emph {et~al.}(2008)\citenamefont
  {Talbayev}, \citenamefont {Trugman}, \citenamefont {Balatsky}, \citenamefont
  {Kimura}, \citenamefont {Taylor},\ and\ \citenamefont
  {Averitt}}]{Talbayev2008}%
  \BibitemOpen
  \bibfield  {author} {\bibinfo {author} {\bibfnamefont {D.}~\bibnamefont
  {Talbayev}}, \bibinfo {author} {\bibfnamefont {S.~A.}\ \bibnamefont
  {Trugman}}, \bibinfo {author} {\bibfnamefont {A.~V.}\ \bibnamefont
  {Balatsky}}, \bibinfo {author} {\bibfnamefont {T.}~\bibnamefont {Kimura}},
  \bibinfo {author} {\bibfnamefont {A.~J.}\ \bibnamefont {Taylor}}, \ and\
  \bibinfo {author} {\bibfnamefont {R.~D.}\ \bibnamefont {Averitt}},\ }\href
  {\doibase 10.1103/PhysRevLett.101.097603} {\bibfield  {journal} {\bibinfo
  {journal} {Physical Review Letters}\ }\textbf {\bibinfo {volume} {101}},\
  \bibinfo {pages} {097603} (\bibinfo {year} {2008})},\ \Eprint
  {http://arxiv.org/abs/0809.0672} {arXiv:0809.0672} \BibitemShut {NoStop}%
\bibitem [{\citenamefont {Kalashnikova}\ \emph {et~al.}(2007)\citenamefont
  {Kalashnikova}, \citenamefont {Kimel}, \citenamefont {Pisarev}, \citenamefont
  {Gridnev}, \citenamefont {Kirilyuk},\ and\ \citenamefont
  {Rasing}}]{Kalashnikova2007}%
  \BibitemOpen
  \bibfield  {author} {\bibinfo {author} {\bibfnamefont {A.~M.}\ \bibnamefont
  {Kalashnikova}}, \bibinfo {author} {\bibfnamefont {A.~V.}\ \bibnamefont
  {Kimel}}, \bibinfo {author} {\bibfnamefont {R.~V.}\ \bibnamefont {Pisarev}},
  \bibinfo {author} {\bibfnamefont {V.~N.}\ \bibnamefont {Gridnev}}, \bibinfo
  {author} {\bibfnamefont {A.}~\bibnamefont {Kirilyuk}}, \ and\ \bibinfo
  {author} {\bibfnamefont {T.}~\bibnamefont {Rasing}},\ }\href@noop {}
  {\bibfield  {journal} {\bibinfo  {journal} {Physical review letters}\
  }\textbf {\bibinfo {volume} {99}},\ \bibinfo {pages} {167205} (\bibinfo
  {year} {2007})}\BibitemShut {NoStop}%
\bibitem [{\citenamefont {Kampfrath}\ \emph {et~al.}(2011)\citenamefont
  {Kampfrath}, \citenamefont {Sell}, \citenamefont {Klatt}, \citenamefont
  {Pashkin}, \citenamefont {M{\"{a}}hrlein}, \citenamefont {Dekorsy},
  \citenamefont {Wolf}, \citenamefont {Fiebig}, \citenamefont {Leitenstorfer},\
  and\ \citenamefont {Huber}}]{Kampfrath2011}%
  \BibitemOpen
  \bibfield  {author} {\bibinfo {author} {\bibfnamefont {T.}~\bibnamefont
  {Kampfrath}}, \bibinfo {author} {\bibfnamefont {A.}~\bibnamefont {Sell}},
  \bibinfo {author} {\bibfnamefont {G.}~\bibnamefont {Klatt}}, \bibinfo
  {author} {\bibfnamefont {A.}~\bibnamefont {Pashkin}}, \bibinfo {author}
  {\bibfnamefont {S.}~\bibnamefont {M{\"{a}}hrlein}}, \bibinfo {author}
  {\bibfnamefont {T.}~\bibnamefont {Dekorsy}}, \bibinfo {author} {\bibfnamefont
  {M.}~\bibnamefont {Wolf}}, \bibinfo {author} {\bibfnamefont {M.}~\bibnamefont
  {Fiebig}}, \bibinfo {author} {\bibfnamefont {A.}~\bibnamefont
  {Leitenstorfer}}, \ and\ \bibinfo {author} {\bibfnamefont {R.}~\bibnamefont
  {Huber}},\ }\href@noop {} {\bibfield  {journal} {\bibinfo  {journal} {Nature
  Photonics}\ }\textbf {\bibinfo {volume} {5}},\ \bibinfo {pages} {31}
  (\bibinfo {year} {2011})}\BibitemShut {NoStop}%
\bibitem [{\citenamefont {Yamaguchi}\ \emph {et~al.}(2010)\citenamefont
  {Yamaguchi}, \citenamefont {Nakajima},\ and\ \citenamefont
  {Suemoto}}]{Yamaguchi2010a}%
  \BibitemOpen
  \bibfield  {author} {\bibinfo {author} {\bibfnamefont {K.}~\bibnamefont
  {Yamaguchi}}, \bibinfo {author} {\bibfnamefont {M.}~\bibnamefont {Nakajima}},
  \ and\ \bibinfo {author} {\bibfnamefont {T.}~\bibnamefont {Suemoto}},\ }\href
  {\doibase 10.1103/PhysRevLett.105.237201} {\bibfield  {journal} {\bibinfo
  {journal} {Physical Review Letters}\ }\textbf {\bibinfo {volume} {105}},\
  \bibinfo {pages} {237201} (\bibinfo {year} {2010})}\BibitemShut {NoStop}%
\bibitem [{\citenamefont {Tokura}\ \emph {et~al.}(2014)\citenamefont {Tokura},
  \citenamefont {Seki},\ and\ \citenamefont {Nagaosa}}]{Tokura2014}%
  \BibitemOpen
  \bibfield  {author} {\bibinfo {author} {\bibfnamefont {Y.}~\bibnamefont
  {Tokura}}, \bibinfo {author} {\bibfnamefont {S.}~\bibnamefont {Seki}}, \ and\
  \bibinfo {author} {\bibfnamefont {N.}~\bibnamefont {Nagaosa}},\ }\href@noop
  {} {\bibfield  {journal} {\bibinfo  {journal} {Reports on Progress in
  Physics}\ }\textbf {\bibinfo {volume} {77}},\ \bibinfo {pages} {76501}
  (\bibinfo {year} {2014})}\BibitemShut {NoStop}%
\bibitem [{\citenamefont {Sivarajah}\ \emph {et~al.}(2017)\citenamefont
  {Sivarajah}, \citenamefont {Steinbacher}, \citenamefont {Dastrup},\ and\
  \citenamefont {Nelson}}]{Sivarajah2017}%
  \BibitemOpen
  \bibfield  {author} {\bibinfo {author} {\bibfnamefont {P.}~\bibnamefont
  {Sivarajah}}, \bibinfo {author} {\bibfnamefont {A.}~\bibnamefont
  {Steinbacher}}, \bibinfo {author} {\bibfnamefont {B.}~\bibnamefont
  {Dastrup}}, \ and\ \bibinfo {author} {\bibfnamefont {K.}~\bibnamefont
  {Nelson}},\ }\href@noop {} {\bibfield  {journal} {\bibinfo  {journal} {arXiv
  preprint arXiv:1707.03503}\ } (\bibinfo {year} {2017})}\BibitemShut {NoStop}%
\bibitem [{\citenamefont {Feurer}\ \emph {et~al.}(2007)\citenamefont {Feurer},
  \citenamefont {Stoyanov}, \citenamefont {Ward}, \citenamefont {Vaughan},
  \citenamefont {Statz},\ and\ \citenamefont {Nelson}}]{Feurer2007ARM}%
  \BibitemOpen
  \bibfield  {author} {\bibinfo {author} {\bibfnamefont {T.}~\bibnamefont
  {Feurer}}, \bibinfo {author} {\bibfnamefont {N.~S.}\ \bibnamefont
  {Stoyanov}}, \bibinfo {author} {\bibfnamefont {D.~W.}\ \bibnamefont {Ward}},
  \bibinfo {author} {\bibfnamefont {J.~C.}\ \bibnamefont {Vaughan}}, \bibinfo
  {author} {\bibfnamefont {E.~R.}\ \bibnamefont {Statz}}, \ and\ \bibinfo
  {author} {\bibfnamefont {K.~A.}\ \bibnamefont {Nelson}},\ }\href@noop {}
  {\bibfield  {journal} {\bibinfo  {journal} {Annual Review of Materials
  Research}\ }\textbf {\bibinfo {volume} {37}},\ \bibinfo {pages} {317}
  (\bibinfo {year} {2007})}\BibitemShut {NoStop}%
\bibitem [{\citenamefont {Feurer}\ \emph {et~al.}(2003)\citenamefont {Feurer},
  \citenamefont {Vaughan},\ and\ \citenamefont {Nelson}}]{Feurer2003c}%
  \BibitemOpen
  \bibfield  {author} {\bibinfo {author} {\bibfnamefont {T.}~\bibnamefont
  {Feurer}}, \bibinfo {author} {\bibfnamefont {J.~C.}\ \bibnamefont {Vaughan}},
  \ and\ \bibinfo {author} {\bibfnamefont {K.~A.}\ \bibnamefont {Nelson}},\
  }\href@noop {} {\bibfield  {journal} {\bibinfo  {journal} {Science (New York,
  N.Y.)}\ }\textbf {\bibinfo {volume} {299}},\ \bibinfo {pages} {374} (\bibinfo
  {year} {2003})}\BibitemShut {NoStop}%
\bibitem [{\citenamefont {Cheung}\ and\ \citenamefont
  {Auston}(1985)}]{Cheung1985}%
  \BibitemOpen
  \bibfield  {author} {\bibinfo {author} {\bibfnamefont {K.~P.}\ \bibnamefont
  {Cheung}}\ and\ \bibinfo {author} {\bibfnamefont {D.~H.}\ \bibnamefont
  {Auston}},\ }\href@noop {} {\bibfield  {journal} {\bibinfo  {journal}
  {Physical review letters}\ }\textbf {\bibinfo {volume} {55}},\ \bibinfo
  {pages} {2152} (\bibinfo {year} {1985})}\BibitemShut {NoStop}%
\bibitem [{\citenamefont {Kaminow}\ and\ \citenamefont {{Johnston
  Jr}}(1967)}]{Kaminow1967}%
  \BibitemOpen
  \bibfield  {author} {\bibinfo {author} {\bibfnamefont {I.~P.}\ \bibnamefont
  {Kaminow}}\ and\ \bibinfo {author} {\bibfnamefont {W.~D.}\ \bibnamefont
  {{Johnston Jr}}},\ }\href@noop {} {\bibfield  {journal} {\bibinfo  {journal}
  {Physical Review}\ }\textbf {\bibinfo {volume} {160}},\ \bibinfo {pages}
  {519} (\bibinfo {year} {1967})}\BibitemShut {NoStop}%
\bibitem [{\citenamefont {Shapiro}\ \emph {et~al.}(1974)\citenamefont
  {Shapiro}, \citenamefont {Axe},\ and\ \citenamefont {Remeika}}]{Shapiro1974}%
  \BibitemOpen
  \bibfield  {author} {\bibinfo {author} {\bibfnamefont {S.~M.}\ \bibnamefont
  {Shapiro}}, \bibinfo {author} {\bibfnamefont {J.~D.}\ \bibnamefont {Axe}}, \
  and\ \bibinfo {author} {\bibfnamefont {J.~P.}\ \bibnamefont {Remeika}},\
  }\href@noop {} {\bibfield  {journal} {\bibinfo  {journal} {Physical Review
  B}\ }\textbf {\bibinfo {volume} {10}},\ \bibinfo {pages} {2014} (\bibinfo
  {year} {1974})}\BibitemShut {NoStop}%
\bibitem [{\citenamefont {Sanders}\ \emph {et~al.}(1978)\citenamefont
  {Sanders}, \citenamefont {Jaccarino},\ and\ \citenamefont
  {Rezende}}]{Sanders1978}%
  \BibitemOpen
  \bibfield  {author} {\bibinfo {author} {\bibfnamefont {R.~W.}\ \bibnamefont
  {Sanders}}, \bibinfo {author} {\bibfnamefont {V.}~\bibnamefont {Jaccarino}},
  \ and\ \bibinfo {author} {\bibfnamefont {S.~M.}\ \bibnamefont {Rezende}},\
  }\href@noop {} {\bibfield  {journal} {\bibinfo  {journal} {Solid State
  Communications}\ }\textbf {\bibinfo {volume} {28}},\ \bibinfo {pages} {907}
  (\bibinfo {year} {1978})}\BibitemShut {NoStop}%
\bibitem [{\citenamefont {Dougherty}\ \emph {et~al.}(1992)\citenamefont
  {Dougherty}, \citenamefont {Wiederrecht},\ and\ \citenamefont
  {Nelson}}]{Dougherty1992a}%
  \BibitemOpen
  \bibfield  {author} {\bibinfo {author} {\bibfnamefont {T.}~\bibnamefont
  {Dougherty}}, \bibinfo {author} {\bibfnamefont {G.}~\bibnamefont
  {Wiederrecht}}, \ and\ \bibinfo {author} {\bibfnamefont {K.}~\bibnamefont
  {Nelson}},\ }\href {\doibase Doi 10.1364/Josab.9.002179} {\bibfield
  {journal} {\bibinfo  {journal} {JOSA B}\ }\textbf {\bibinfo {volume} {9}},\
  \bibinfo {pages} {2179} (\bibinfo {year} {1992})}\BibitemShut {NoStop}%
\bibitem [{\citenamefont {Yeh}\ \emph {et~al.}(2007{\natexlab{a}})\citenamefont
  {Yeh}, \citenamefont {Hoffmann}, \citenamefont {Hebling},\ and\ \citenamefont
  {Nelson}}]{Yeh2007a}%
  \BibitemOpen
  \bibfield  {author} {\bibinfo {author} {\bibfnamefont {K.-L.}\ \bibnamefont
  {Yeh}}, \bibinfo {author} {\bibfnamefont {M.~C.}\ \bibnamefont {Hoffmann}},
  \bibinfo {author} {\bibfnamefont {J.}~\bibnamefont {Hebling}}, \ and\
  \bibinfo {author} {\bibfnamefont {K.~A.}\ \bibnamefont {Nelson}},\
  }\href@noop {} {\bibfield  {journal} {\bibinfo  {journal} {Appl. Phys.
  Lett.}\ }\textbf {\bibinfo {volume} {90}},\ \bibinfo {pages} {171121}
  (\bibinfo {year} {2007}{\natexlab{a}})}\BibitemShut {NoStop}%
\bibitem [{\citenamefont {Yang}\ \emph {et~al.}(2010)\citenamefont {Yang},
  \citenamefont {Wu}, \citenamefont {Xu}, \citenamefont {Nelson},\ and\
  \citenamefont {Werley}}]{Yang2010}%
  \BibitemOpen
  \bibfield  {author} {\bibinfo {author} {\bibfnamefont {C.}~\bibnamefont
  {Yang}}, \bibinfo {author} {\bibfnamefont {Q.}~\bibnamefont {Wu}}, \bibinfo
  {author} {\bibfnamefont {J.}~\bibnamefont {Xu}}, \bibinfo {author}
  {\bibfnamefont {K.~A.}\ \bibnamefont {Nelson}}, \ and\ \bibinfo {author}
  {\bibfnamefont {C.~A.}\ \bibnamefont {Werley}},\ }\href@noop {} {\bibfield
  {journal} {\bibinfo  {journal} {Opt. Express}\ }\textbf {\bibinfo {volume}
  {18}},\ \bibinfo {pages} {26351} (\bibinfo {year} {2010})}\BibitemShut
  {NoStop}%
\bibitem [{\citenamefont {Winnewisser}\ \emph {et~al.}(1997)\citenamefont
  {Winnewisser}, \citenamefont {Jepsen}, \citenamefont {Schall}, \citenamefont
  {Schyja},\ and\ \citenamefont {Helm}}]{Winnewisser1997a}%
  \BibitemOpen
  \bibfield  {author} {\bibinfo {author} {\bibfnamefont {C.}~\bibnamefont
  {Winnewisser}}, \bibinfo {author} {\bibfnamefont {P.~U.}\ \bibnamefont
  {Jepsen}}, \bibinfo {author} {\bibfnamefont {M.}~\bibnamefont {Schall}},
  \bibinfo {author} {\bibfnamefont {V.}~\bibnamefont {Schyja}}, \ and\ \bibinfo
  {author} {\bibfnamefont {H.}~\bibnamefont {Helm}},\ }\href {\doibase
  10.1063/1.119093} {\enquote {\bibinfo {title} {{Electro-optic detection of
  THz radiation in LiTaO3, LiNbO3, and ZnTe}},}\ } (\bibinfo {year}
  {1997})\BibitemShut {NoStop}%
\bibitem [{\citenamefont {Pálfalvi}\ \emph {et~al.}(2005)\citenamefont
  {Pálfalvi}, \citenamefont {Hebling}, \citenamefont {Kuhl}, \citenamefont
  {Péter},\ and\ \citenamefont {Polgár}}]{Palfalvi2005b}%
  \BibitemOpen
  \bibfield  {author} {\bibinfo {author} {\bibfnamefont {L.}~\bibnamefont
  {Pálfalvi}}, \bibinfo {author} {\bibfnamefont {J.}~\bibnamefont {Hebling}},
  \bibinfo {author} {\bibfnamefont {J.}~\bibnamefont {Kuhl}}, \bibinfo {author}
  {\bibfnamefont {A.}~\bibnamefont {Péter}}, \ and\ \bibinfo {author}
  {\bibfnamefont {K.}~\bibnamefont {Polgár}},\ }\href {\doibase
  10.1063/1.1929859} {\bibfield  {journal} {\bibinfo  {journal} {Journal of
  Applied Physics}\ }\textbf {\bibinfo {volume} {97}},\ \bibinfo {pages} {3505}
  (\bibinfo {year} {2005})}\BibitemShut {NoStop}%
\bibitem [{\citenamefont {Johnson}\ and\ \citenamefont
  {Joannopoulos}(2001)}]{Johnson2001}%
  \BibitemOpen
  \bibfield  {author} {\bibinfo {author} {\bibfnamefont {S.~G.}\ \bibnamefont
  {Johnson}}\ and\ \bibinfo {author} {\bibfnamefont {J.~D.}\ \bibnamefont
  {Joannopoulos}},\ }\href@noop {} {\bibfield  {journal} {\bibinfo  {journal}
  {Optics express}\ }\textbf {\bibinfo {volume} {8}},\ \bibinfo {pages} {173}
  (\bibinfo {year} {2001})}\BibitemShut {NoStop}%
\bibitem [{\citenamefont {Kozlov}\ \emph {et~al.}(1993)\citenamefont {Kozlov},
  \citenamefont {Lebedev}, \citenamefont {Mukhin}, \citenamefont {Prokhorov},
  \citenamefont {Fedorov}, \citenamefont {Balbashov},\ and\ \citenamefont
  {Parsegov}}]{Kozlov1993}%
  \BibitemOpen
  \bibfield  {author} {\bibinfo {author} {\bibfnamefont {G.~V.}\ \bibnamefont
  {Kozlov}}, \bibinfo {author} {\bibfnamefont {S.~P.}\ \bibnamefont {Lebedev}},
  \bibinfo {author} {\bibfnamefont {A.~A.}\ \bibnamefont {Mukhin}}, \bibinfo
  {author} {\bibfnamefont {A.~S.}\ \bibnamefont {Prokhorov}}, \bibinfo {author}
  {\bibfnamefont {L.~V.}\ \bibnamefont {Fedorov}}, \bibinfo {author}
  {\bibfnamefont {A.~M.}\ \bibnamefont {Balbashov}}, \ and\ \bibinfo {author}
  {\bibfnamefont {I.~Y.}\ \bibnamefont {Parsegov}},\ }\href@noop {} {\bibfield
  {journal} {\bibinfo  {journal} {Magnetics, IEEE Transactions on}\ }\textbf
  {\bibinfo {volume} {29}},\ \bibinfo {pages} {3443} (\bibinfo {year}
  {1993})}\BibitemShut {NoStop}%
\bibitem [{\citenamefont {Zhu}\ \emph {et~al.}(1990)\citenamefont {Zhu},
  \citenamefont {Gauthier}, \citenamefont {Morin}, \citenamefont {Wu},
  \citenamefont {Carmichael},\ and\ \citenamefont {Mossberg}}]{Zhu1990}%
  \BibitemOpen
  \bibfield  {author} {\bibinfo {author} {\bibfnamefont {Y.}~\bibnamefont
  {Zhu}}, \bibinfo {author} {\bibfnamefont {D.~J.}\ \bibnamefont {Gauthier}},
  \bibinfo {author} {\bibfnamefont {S.~E.}\ \bibnamefont {Morin}}, \bibinfo
  {author} {\bibfnamefont {Q.}~\bibnamefont {Wu}}, \bibinfo {author}
  {\bibfnamefont {H.~J.}\ \bibnamefont {Carmichael}}, \ and\ \bibinfo {author}
  {\bibfnamefont {T.~W.}\ \bibnamefont {Mossberg}},\ }\href {\doibase
  10.1103/PhysRevLett.64.2499} {\bibfield  {journal} {\bibinfo  {journal}
  {Physical Review Letters}\ }\textbf {\bibinfo {volume} {64}},\ \bibinfo
  {pages} {2499} (\bibinfo {year} {1990})}\BibitemShut {NoStop}%
\bibitem [{\citenamefont {Fink}\ \emph {et~al.}(2010)\citenamefont {Fink},
  \citenamefont {Steffen}, \citenamefont {Studer}, \citenamefont {Bishop},
  \citenamefont {Baur}, \citenamefont {Bianchetti}, \citenamefont {Bozyigit},
  \citenamefont {Lang}, \citenamefont {Filipp}, \citenamefont {Leek},\ and\
  \citenamefont {Wallraff}}]{Fink2010}%
  \BibitemOpen
  \bibfield  {author} {\bibinfo {author} {\bibfnamefont {J.~M.}\ \bibnamefont
  {Fink}}, \bibinfo {author} {\bibfnamefont {L.}~\bibnamefont {Steffen}},
  \bibinfo {author} {\bibfnamefont {P.}~\bibnamefont {Studer}}, \bibinfo
  {author} {\bibfnamefont {L.~S.}\ \bibnamefont {Bishop}}, \bibinfo {author}
  {\bibfnamefont {M.}~\bibnamefont {Baur}}, \bibinfo {author} {\bibfnamefont
  {R.}~\bibnamefont {Bianchetti}}, \bibinfo {author} {\bibfnamefont
  {D.}~\bibnamefont {Bozyigit}}, \bibinfo {author} {\bibfnamefont
  {C.}~\bibnamefont {Lang}}, \bibinfo {author} {\bibfnamefont {S.}~\bibnamefont
  {Filipp}}, \bibinfo {author} {\bibfnamefont {P.~J.}\ \bibnamefont {Leek}}, \
  and\ \bibinfo {author} {\bibfnamefont {A.}~\bibnamefont {Wallraff}},\ }\href
  {\doibase 10.1103/PhysRevLett.105.163601} {\bibfield  {journal} {\bibinfo
  {journal} {Physical Review Letters}\ }\textbf {\bibinfo {volume} {105}}
  (\bibinfo {year} {2010}),\ 10.1103/PhysRevLett.105.163601},\ \Eprint
  {http://arxiv.org/abs/1003.1161} {arXiv:1003.1161} \BibitemShut {NoStop}%
\bibitem [{\citenamefont {Nova}\ \emph {et~al.}(2016)\citenamefont {Nova},
  \citenamefont {Cartella}, \citenamefont {Cantaluppi}, \citenamefont
  {F{\"{o}}rst}, \citenamefont {Bossini}, \citenamefont {Mikhaylovskiy},
  \citenamefont {Kimel}, \citenamefont {Merlin},\ and\ \citenamefont
  {Cavalleri}}]{Nova2016}%
  \BibitemOpen
  \bibfield  {author} {\bibinfo {author} {\bibfnamefont {T.~F.}\ \bibnamefont
  {Nova}}, \bibinfo {author} {\bibfnamefont {A.}~\bibnamefont {Cartella}},
  \bibinfo {author} {\bibfnamefont {A.}~\bibnamefont {Cantaluppi}}, \bibinfo
  {author} {\bibfnamefont {M.}~\bibnamefont {F{\"{o}}rst}}, \bibinfo {author}
  {\bibfnamefont {D.}~\bibnamefont {Bossini}}, \bibinfo {author} {\bibfnamefont
  {R.~V.}\ \bibnamefont {Mikhaylovskiy}}, \bibinfo {author} {\bibfnamefont
  {A.~V.}\ \bibnamefont {Kimel}}, \bibinfo {author} {\bibfnamefont
  {R.}~\bibnamefont {Merlin}}, \ and\ \bibinfo {author} {\bibfnamefont
  {A.}~\bibnamefont {Cavalleri}},\ }\href@noop {} {\bibfield  {journal}
  {\bibinfo  {journal} {Nature Physics}\ }\textbf {\bibinfo {volume} {1}},\
  \bibinfo {pages} {1} (\bibinfo {year} {2016})}\BibitemShut {NoStop}%
\bibitem [{\citenamefont {Zhou}\ \emph {et~al.}(2007)\citenamefont {Zhou},
  \citenamefont {Koschny},\ and\ \citenamefont {Soukoulis}}]{Zhou2007}%
  \BibitemOpen
  \bibfield  {author} {\bibinfo {author} {\bibfnamefont {J.}~\bibnamefont
  {Zhou}}, \bibinfo {author} {\bibfnamefont {T.}~\bibnamefont {Koschny}}, \
  and\ \bibinfo {author} {\bibfnamefont {C.~M.}\ \bibnamefont {Soukoulis}},\
  }\href {\doibase 10.1364/OE.15.017881} {\bibfield  {journal} {\bibinfo
  {journal} {Optics express}\ }\textbf {\bibinfo {volume} {15}},\ \bibinfo
  {pages} {17881} (\bibinfo {year} {2007})},\ \Eprint
  {http://arxiv.org/abs/0710.0812} {arXiv:0710.0812} \BibitemShut {NoStop}%
\bibitem [{\citenamefont {Joannopoulos}\ \emph {et~al.}(2008)\citenamefont
  {Joannopoulos}, \citenamefont {Johnson}, \citenamefont {Winn},\ and\
  \citenamefont {Meade}}]{Joannopoulos2008}%
  \BibitemOpen
  \bibfield  {author} {\bibinfo {author} {\bibfnamefont {J.~D.}\ \bibnamefont
  {Joannopoulos}}, \bibinfo {author} {\bibfnamefont {S.~G.}\ \bibnamefont
  {Johnson}}, \bibinfo {author} {\bibfnamefont {J.~N.}\ \bibnamefont {Winn}}, \
  and\ \bibinfo {author} {\bibfnamefont {R.~D.}\ \bibnamefont {Meade}},\
  }\href@noop {} {\emph {\bibinfo {title} {{Photonic Crystals Molding the Flow
  of Light}}}},\ \bibinfo {edition} {2nd}\ ed.\ (\bibinfo  {publisher}
  {Princeton University Press, Princeton},\ \bibinfo {address} {Princeton, New
  Jersey},\ \bibinfo {year} {2008})\ pp.\ \bibinfo {pages} {52--54}\BibitemShut
  {NoStop}%
\bibitem [{Note1()}]{Note1}%
  \BibitemOpen
  \bibinfo {note} {Note that the space-time plot in Fig.~\ref {fig:setup}b,
  which was used for the eigenmode analysis starts, at approximately 0.5 mm
  into the hybrid structure in order to avoid significant evanescent wave
  contributions~\cite {Feurer2007ARM}}\BibitemShut {NoStop}%
\bibitem [{\citenamefont {Cho}\ and\ \citenamefont
  {Yamanouchi}(1987)}]{Cho1987}%
  \BibitemOpen
  \bibfield  {author} {\bibinfo {author} {\bibfnamefont {Y.}~\bibnamefont
  {Cho}}\ and\ \bibinfo {author} {\bibfnamefont {K.}~\bibnamefont
  {Yamanouchi}},\ }\href {\doibase 10.1063/1.338138} {\bibfield  {journal}
  {\bibinfo  {journal} {Journal of Applied Physics}\ }\textbf {\bibinfo
  {volume} {61}},\ \bibinfo {pages} {875} (\bibinfo {year} {1987})}\BibitemShut
  {NoStop}%
\bibitem [{\citenamefont {Huang}(1951)}]{Huang1951a}%
  \BibitemOpen
  \bibfield  {author} {\bibinfo {author} {\bibfnamefont {K.}~\bibnamefont
  {Huang}},\ }in\ \href@noop {} {\emph {\bibinfo {booktitle} {Proceedings of
  the Royal Society of London A: Mathematical, Physical and Engineering
  Sciences}}},\ Vol.\ \bibinfo {volume} {208}\ (\bibinfo  {publisher} {The
  Royal Society},\ \bibinfo {year} {1951})\ pp.\ \bibinfo {pages}
  {352--365}\BibitemShut {NoStop}%
\bibitem [{\citenamefont {Lee}\ \emph {et~al.}(2001)\citenamefont {Lee},
  \citenamefont {Meade}, \citenamefont {Norris},\ and\ \citenamefont
  {Galvanauskas}}]{Lee2001}%
  \BibitemOpen
  \bibfield  {author} {\bibinfo {author} {\bibfnamefont {Y.~S.}\ \bibnamefont
  {Lee}}, \bibinfo {author} {\bibfnamefont {T.}~\bibnamefont {Meade}}, \bibinfo
  {author} {\bibfnamefont {T.~B.}\ \bibnamefont {Norris}}, \ and\ \bibinfo
  {author} {\bibfnamefont {A.}~\bibnamefont {Galvanauskas}},\ }\href {\doibase
  10.1063/1.1373406} {\bibfield  {journal} {\bibinfo  {journal} {Applied
  Physics Letters}\ }\textbf {\bibinfo {volume} {78}},\ \bibinfo {pages} {3583}
  (\bibinfo {year} {2001})}\BibitemShut {NoStop}%
\bibitem [{\citenamefont {Nishitani}\ \emph {et~al.}(2010)\citenamefont
  {Nishitani}, \citenamefont {Nagashima},\ and\ \citenamefont
  {Hangyo}}]{Nishitani2010}%
  \BibitemOpen
  \bibfield  {author} {\bibinfo {author} {\bibfnamefont {J.}~\bibnamefont
  {Nishitani}}, \bibinfo {author} {\bibfnamefont {T.}~\bibnamefont
  {Nagashima}}, \ and\ \bibinfo {author} {\bibfnamefont {M.}~\bibnamefont
  {Hangyo}},\ }in\ \href {\doibase 10.1109/ICIMW.2010.5612484} {\emph {\bibinfo
  {booktitle} {IRMMW-THz 2010 - 35th International Conference on Infrared,
  Millimeter, and Terahertz Waves, Conference Guide}}}\ (\bibinfo {year}
  {2010})\BibitemShut {NoStop}%
\bibitem [{\citenamefont {Fujiwara}\ and\ \citenamefont
  {Wakatsuki}(1987)}]{Fujiwara1987}%
  \BibitemOpen
  \bibfield  {author} {\bibinfo {author} {\bibfnamefont {Y.}~\bibnamefont
  {Fujiwara}}\ and\ \bibinfo {author} {\bibfnamefont {N.}~\bibnamefont
  {Wakatsuki}},\ }\href@noop {} {\bibfield  {journal} {\bibinfo  {journal}
  {IEEE Transactions on Ultrasonics Ferroelectrics and Frequency Control}\
  }\textbf {\bibinfo {volume} {34}},\ \bibinfo {pages} {39} (\bibinfo {year}
  {1987})}\BibitemShut {NoStop}%
\bibitem [{\citenamefont {Laurence}\ and\ \citenamefont
  {Petitgrand}(1973)}]{Laurence1973}%
  \BibitemOpen
  \bibfield  {author} {\bibinfo {author} {\bibfnamefont {G.}~\bibnamefont
  {Laurence}}\ and\ \bibinfo {author} {\bibfnamefont {D.}~\bibnamefont
  {Petitgrand}},\ }\href@noop {} {\bibfield  {journal} {\bibinfo  {journal}
  {Physical Review B}\ }\textbf {\bibinfo {volume} {8}},\ \bibinfo {pages}
  {2130} (\bibinfo {year} {1973})}\BibitemShut {NoStop}%
\bibitem [{\citenamefont {Ando}\ \emph {et~al.}(2011)\citenamefont {Ando},
  \citenamefont {Takahashi}, \citenamefont {Ieda}, \citenamefont {Kurebayashi},
  \citenamefont {Trypiniotis}, \citenamefont {Barnes}, \citenamefont
  {Maekawa},\ and\ \citenamefont {Saitoh}}]{Ando2011}%
  \BibitemOpen
  \bibfield  {author} {\bibinfo {author} {\bibfnamefont {K.}~\bibnamefont
  {Ando}}, \bibinfo {author} {\bibfnamefont {S.}~\bibnamefont {Takahashi}},
  \bibinfo {author} {\bibfnamefont {J.}~\bibnamefont {Ieda}}, \bibinfo {author}
  {\bibfnamefont {H.}~\bibnamefont {Kurebayashi}}, \bibinfo {author}
  {\bibfnamefont {T.}~\bibnamefont {Trypiniotis}}, \bibinfo {author}
  {\bibfnamefont {C.~H.~W.}\ \bibnamefont {Barnes}}, \bibinfo {author}
  {\bibfnamefont {S.}~\bibnamefont {Maekawa}}, \ and\ \bibinfo {author}
  {\bibfnamefont {E.}~\bibnamefont {Saitoh}},\ }\href {\doibase
  10.1038/nmat3052} {\bibfield  {journal} {\bibinfo  {journal} {Nature
  Materials}\ }\textbf {\bibinfo {volume} {10}},\ \bibinfo {pages} {655}
  (\bibinfo {year} {2011})}\BibitemShut {NoStop}%
\bibitem [{\citenamefont {Novotny}(2010)}]{Novotny2010}%
  \BibitemOpen
  \bibfield  {author} {\bibinfo {author} {\bibfnamefont {L.}~\bibnamefont
  {Novotny}},\ }\href {\doibase 10.1119/1.3471177} {\bibfield  {journal}
  {\bibinfo  {journal} {American Journal of Physics}\ }\textbf {\bibinfo
  {volume} {78}},\ \bibinfo {pages} {1199} (\bibinfo {year}
  {2010})}\BibitemShut {NoStop}%
\bibitem [{\citenamefont {Schuster}\ \emph {et~al.}(2010)\citenamefont
  {Schuster}, \citenamefont {Sears}, \citenamefont {Ginossar}, \citenamefont
  {Dicarlo}, \citenamefont {Frunzio}, \citenamefont {Morton}, \citenamefont
  {Wu}, \citenamefont {Briggs}, \citenamefont {Buckley}, \citenamefont
  {Awschalom},\ and\ \citenamefont {Schoelkopf}}]{Schuster2010}%
  \BibitemOpen
  \bibfield  {author} {\bibinfo {author} {\bibfnamefont {D.~I.}\ \bibnamefont
  {Schuster}}, \bibinfo {author} {\bibfnamefont {A.~P.}\ \bibnamefont {Sears}},
  \bibinfo {author} {\bibfnamefont {E.}~\bibnamefont {Ginossar}}, \bibinfo
  {author} {\bibfnamefont {L.}~\bibnamefont {Dicarlo}}, \bibinfo {author}
  {\bibfnamefont {L.}~\bibnamefont {Frunzio}}, \bibinfo {author} {\bibfnamefont
  {J.~J.~L.}\ \bibnamefont {Morton}}, \bibinfo {author} {\bibfnamefont
  {H.}~\bibnamefont {Wu}}, \bibinfo {author} {\bibfnamefont {G.~A.~D.}\
  \bibnamefont {Briggs}}, \bibinfo {author} {\bibfnamefont {B.~B.}\
  \bibnamefont {Buckley}}, \bibinfo {author} {\bibfnamefont {D.~D.}\
  \bibnamefont {Awschalom}}, \ and\ \bibinfo {author} {\bibfnamefont {R.~J.}\
  \bibnamefont {Schoelkopf}},\ }\href {\doibase 10.1103/PhysRevLett.105.140501}
  {\bibfield  {journal} {\bibinfo  {journal} {Physical Review Letters}\
  }\textbf {\bibinfo {volume} {105}} (\bibinfo {year} {2010}),\
  10.1103/PhysRevLett.105.140501},\ \Eprint {http://arxiv.org/abs/1006.0242}
  {arXiv:1006.0242} \BibitemShut {NoStop}%
\bibitem [{\citenamefont {Wolf}\ \emph {et~al.}(2001)\citenamefont {Wolf},
  \citenamefont {Awschalom}, \citenamefont {Buhrman}, \citenamefont {Daughton},
  \citenamefont {Moln{\'{a}}r}, \citenamefont {Roukes}, \citenamefont
  {Chtchelkanova},\ and\ \citenamefont {Treger}}]{Wolf2016a}%
  \BibitemOpen
  \bibfield  {author} {\bibinfo {author} {\bibfnamefont {A.~S.~A.}\
  \bibnamefont {Wolf}}, \bibinfo {author} {\bibfnamefont {D.~D.}\ \bibnamefont
  {Awschalom}}, \bibinfo {author} {\bibfnamefont {R.~A.}\ \bibnamefont
  {Buhrman}}, \bibinfo {author} {\bibfnamefont {J.~M.}\ \bibnamefont
  {Daughton}}, \bibinfo {author} {\bibfnamefont {S.~V.}\ \bibnamefont
  {Moln{\'{a}}r}}, \bibinfo {author} {\bibfnamefont {L.}~\bibnamefont
  {Roukes}}, \bibinfo {author} {\bibfnamefont {A.~Y.}\ \bibnamefont
  {Chtchelkanova}}, \ and\ \bibinfo {author} {\bibfnamefont {D.~M.}\
  \bibnamefont {Treger}},\ }\href@noop {} {\ \textbf {\bibinfo {volume}
  {294}},\ \bibinfo {pages} {1488} (\bibinfo {year} {2001})}\BibitemShut
  {NoStop}%
\bibitem [{\citenamefont {Weisbuch}\ \emph {et~al.}(1992)\citenamefont
  {Weisbuch}, \citenamefont {Nishioka}, \citenamefont {Ishikawa},\ and\
  \citenamefont {Arakawa}}]{Weisbuch1992}%
  \BibitemOpen
  \bibfield  {author} {\bibinfo {author} {\bibfnamefont {C.}~\bibnamefont
  {Weisbuch}}, \bibinfo {author} {\bibfnamefont {M.}~\bibnamefont {Nishioka}},
  \bibinfo {author} {\bibfnamefont {A.}~\bibnamefont {Ishikawa}}, \ and\
  \bibinfo {author} {\bibfnamefont {Y.}~\bibnamefont {Arakawa}},\ }\href
  {\doibase 10.1103/PhysRevLett.69.3314} {\bibfield  {journal} {\bibinfo
  {journal} {Physical Review Letters}\ }\textbf {\bibinfo {volume} {69}},\
  \bibinfo {pages} {3314} (\bibinfo {year} {1992})}\BibitemShut {NoStop}%
\bibitem [{\citenamefont {Blais}\ \emph {et~al.}(2004)\citenamefont {Blais},
  \citenamefont {Huang}, \citenamefont {Wallraff}, \citenamefont {Girvin},\
  and\ \citenamefont {Schoelkopf}}]{Blais2004a}%
  \BibitemOpen
  \bibfield  {author} {\bibinfo {author} {\bibfnamefont {A.}~\bibnamefont
  {Blais}}, \bibinfo {author} {\bibfnamefont {R.~S.}\ \bibnamefont {Huang}},
  \bibinfo {author} {\bibfnamefont {A.}~\bibnamefont {Wallraff}}, \bibinfo
  {author} {\bibfnamefont {S.~M.}\ \bibnamefont {Girvin}}, \ and\ \bibinfo
  {author} {\bibfnamefont {R.~J.}\ \bibnamefont {Schoelkopf}},\ }\href@noop {}
  {\bibfield  {journal} {\bibinfo  {journal} {Phys. Rev. A.}\ }\textbf
  {\bibinfo {volume} {69}},\ \bibinfo {pages} {062320} (\bibinfo {year}
  {2004})}\BibitemShut {NoStop}%
\bibitem [{\citenamefont {Cheong}\ and\ \citenamefont
  {Mostovoy}(2007)}]{Cheong2007}%
  \BibitemOpen
  \bibfield  {author} {\bibinfo {author} {\bibfnamefont {S.-W.}\ \bibnamefont
  {Cheong}}\ and\ \bibinfo {author} {\bibfnamefont {M.}~\bibnamefont
  {Mostovoy}},\ }\href {\doibase 10.1038/nmat1804} {\bibfield  {journal}
  {\bibinfo  {journal} {Nature Materials}\ }\textbf {\bibinfo {volume} {6}},\
  \bibinfo {pages} {13} (\bibinfo {year} {2007})}\BibitemShut {NoStop}%
\bibitem [{\citenamefont {Brune}\ \emph {et~al.}(1996)\citenamefont {Brune},
  \citenamefont {Schmidt-Kaler}, \citenamefont {Maali}, \citenamefont {Dreyer},
  \citenamefont {Hagley}, \citenamefont {Raimond},\ and\ \citenamefont
  {Haroche}}]{Brune1996}%
  \BibitemOpen
  \bibfield  {author} {\bibinfo {author} {\bibfnamefont {M.}~\bibnamefont
  {Brune}}, \bibinfo {author} {\bibfnamefont {F.}~\bibnamefont
  {Schmidt-Kaler}}, \bibinfo {author} {\bibfnamefont {A.}~\bibnamefont
  {Maali}}, \bibinfo {author} {\bibfnamefont {J.}~\bibnamefont {Dreyer}},
  \bibinfo {author} {\bibfnamefont {E.}~\bibnamefont {Hagley}}, \bibinfo
  {author} {\bibfnamefont {J.~M.}\ \bibnamefont {Raimond}}, \ and\ \bibinfo
  {author} {\bibfnamefont {S.}~\bibnamefont {Haroche}},\ }\href@noop {}
  {\bibfield  {journal} {\bibinfo  {journal} {Physical Review Letters}\
  }\textbf {\bibinfo {volume} {76}},\ \bibinfo {pages} {1800} (\bibinfo {year}
  {1996})}\BibitemShut {NoStop}%
\bibitem [{\citenamefont {{De Sousa}}\ and\ \citenamefont
  {Moore}(2008)}]{DeSousa2008}%
  \BibitemOpen
  \bibfield  {author} {\bibinfo {author} {\bibfnamefont {R.}~\bibnamefont {{De
  Sousa}}}\ and\ \bibinfo {author} {\bibfnamefont {J.~E.}\ \bibnamefont
  {Moore}},\ }\href {\doibase 10.1103/PhysRevB.77.012406} {\bibfield  {journal}
  {\bibinfo  {journal} {Physical Review B - Condensed Matter and Materials
  Physics}\ }\textbf {\bibinfo {volume} {77}},\ \bibinfo {pages} {3} (\bibinfo
  {year} {2008})},\ \Eprint {http://arxiv.org/abs/0706.1260} {arXiv:0706.1260}
  \BibitemShut {NoStop}%
\bibitem [{\citenamefont {Imamoǧlu}(2009)}]{Imamoglu2009}%
  \BibitemOpen
  \bibfield  {author} {\bibinfo {author} {\bibfnamefont {A.}~\bibnamefont
  {Imamoǧlu}},\ }\href {\doibase 10.1103/PhysRevLett.102.083602} {\bibfield
  {journal} {\bibinfo  {journal} {Physical Review Letters}\ }\textbf {\bibinfo
  {volume} {102}} (\bibinfo {year} {2009}),\ 10.1103/PhysRevLett.102.083602},\
  \Eprint {http://arxiv.org/abs/0809.2909} {arXiv:0809.2909} \BibitemShut
  {NoStop}%
\bibitem [{\citenamefont {Shim}\ \emph {et~al.}(2007)\citenamefont {Shim},
  \citenamefont {Imboden},\ and\ \citenamefont {Mohanty}}]{Shim2007}%
  \BibitemOpen
  \bibfield  {author} {\bibinfo {author} {\bibfnamefont {S.-B.}\ \bibnamefont
  {Shim}}, \bibinfo {author} {\bibfnamefont {M.}~\bibnamefont {Imboden}}, \
  and\ \bibinfo {author} {\bibfnamefont {P.}~\bibnamefont {Mohanty}},\ }\href
  {\doibase 10.1126/science.1137307} {\bibfield  {journal} {\bibinfo  {journal}
  {Science (New York, N.Y.)}\ }\textbf {\bibinfo {volume} {316}},\ \bibinfo
  {pages} {95} (\bibinfo {year} {2007})}\BibitemShut {NoStop}%
\bibitem [{\citenamefont {Ida}(2005)}]{Ida2005}%
  \BibitemOpen
  \bibfield  {author} {\bibinfo {author} {\bibfnamefont {M.}~\bibnamefont
  {Ida}},\ }\href {\doibase 10.1103/PhysRevE.72.036306} {\bibfield  {journal}
  {\bibinfo  {journal} {Physical Review E - Statistical, Nonlinear, and Soft
  Matter Physics}\ }\textbf {\bibinfo {volume} {72}} (\bibinfo {year} {2005}),\
  10.1103/PhysRevE.72.036306}\BibitemShut {NoStop}%
\bibitem [{\citenamefont {Wu}\ and\ \citenamefont {Zhang}(1995)}]{Wu1995}%
  \BibitemOpen
  \bibfield  {author} {\bibinfo {author} {\bibfnamefont {Q.}~\bibnamefont
  {Wu}}\ and\ \bibinfo {author} {\bibfnamefont {X.-C.}\ \bibnamefont {Zhang}},\
  }\href {\doibase 10.1063/1.114909} {\bibfield  {journal} {\bibinfo  {journal}
  {Applied Physics Letters}\ }\textbf {\bibinfo {volume} {67}},\ \bibinfo
  {pages} {3523} (\bibinfo {year} {1995})}\BibitemShut {NoStop}%
\bibitem [{\citenamefont {Yeh}\ \emph {et~al.}(2007{\natexlab{b}})\citenamefont
  {Yeh}, \citenamefont {Hoffmann}, \citenamefont {Hebling},\ and\ \citenamefont
  {Nelson}}]{Yeh2007}%
  \BibitemOpen
  \bibfield  {author} {\bibinfo {author} {\bibfnamefont {K.-L.}\ \bibnamefont
  {Yeh}}, \bibinfo {author} {\bibfnamefont {M.~C.}\ \bibnamefont {Hoffmann}},
  \bibinfo {author} {\bibfnamefont {J.}~\bibnamefont {Hebling}}, \ and\
  \bibinfo {author} {\bibfnamefont {K.~A.}\ \bibnamefont {Nelson}},\ }\href
  {http://ieeexplore.ieee.org/xpl/articleDetails.jsp?arnumber=4826359}
  {\bibfield  {journal} {\bibinfo  {journal} {Applied Physics Letters}\
  }\textbf {\bibinfo {volume} {90}},\ \bibinfo {pages} {171121} (\bibinfo
  {year} {2007}{\natexlab{b}})}\BibitemShut {NoStop}%
\bibitem [{\citenamefont {Bogani}\ and\ \citenamefont
  {Wernsdorfer}(2008)}]{Bogani2008}%
  \BibitemOpen
  \bibfield  {author} {\bibinfo {author} {\bibfnamefont {L.}~\bibnamefont
  {Bogani}}\ and\ \bibinfo {author} {\bibfnamefont {W.}~\bibnamefont
  {Wernsdorfer}},\ }\href {\doibase 10.1038/nmat2133} {\bibfield  {journal}
  {\bibinfo  {journal} {Nature materials}\ }\textbf {\bibinfo {volume} {7}},\
  \bibinfo {pages} {179} (\bibinfo {year} {2008})},\ \Eprint
  {http://arxiv.org/abs/arXiv:1011.1669v3} {arXiv:arXiv:1011.1669v3}
  \BibitemShut {NoStop}%
\bibitem [{\citenamefont {Wallquist}\ \emph {et~al.}(2009)\citenamefont
  {Wallquist}, \citenamefont {Hammerer}, \citenamefont {Rabl}, \citenamefont
  {Lukin},\ and\ \citenamefont {Zoller}}]{Wallquist2009}%
  \BibitemOpen
  \bibfield  {author} {\bibinfo {author} {\bibfnamefont {M.}~\bibnamefont
  {Wallquist}}, \bibinfo {author} {\bibfnamefont {K.}~\bibnamefont {Hammerer}},
  \bibinfo {author} {\bibfnamefont {P.}~\bibnamefont {Rabl}}, \bibinfo {author}
  {\bibfnamefont {M.}~\bibnamefont {Lukin}}, \ and\ \bibinfo {author}
  {\bibfnamefont {P.}~\bibnamefont {Zoller}},\ }\href {\doibase
  10.1088/0031-8949/2009/T137/014001} {\bibfield  {journal} {\bibinfo
  {journal} {Physica Scripta}\ }\textbf {\bibinfo {volume} {T137}},\ \bibinfo
  {pages} {014001} (\bibinfo {year} {2009})},\ \Eprint
  {http://arxiv.org/abs/0911.3835} {arXiv:0911.3835} \BibitemShut {NoStop}%
\bibitem [{\citenamefont {Wu}\ \emph {et~al.}(2009)\citenamefont {Wu},
  \citenamefont {Werley}, \citenamefont {Lin}, \citenamefont {Dorn},
  \citenamefont {Bawendi},\ and\ \citenamefont {Nelson}}]{Wu2009b}%
  \BibitemOpen
  \bibfield  {author} {\bibinfo {author} {\bibfnamefont {Q.}~\bibnamefont
  {Wu}}, \bibinfo {author} {\bibfnamefont {C.~A.}\ \bibnamefont {Werley}},
  \bibinfo {author} {\bibfnamefont {K.-H.}\ \bibnamefont {Lin}}, \bibinfo
  {author} {\bibfnamefont {A.}~\bibnamefont {Dorn}}, \bibinfo {author}
  {\bibfnamefont {M.~G.}\ \bibnamefont {Bawendi}}, \ and\ \bibinfo {author}
  {\bibfnamefont {K.~A.}\ \bibnamefont {Nelson}},\ }\href {\doibase
  10.1364/OE.17.009219} {\bibfield  {journal} {\bibinfo  {journal} {Optics
  Express}\ }\textbf {\bibinfo {volume} {17}},\ \bibinfo {pages} {9219}
  (\bibinfo {year} {2009})}\BibitemShut {NoStop}%
\bibitem [{\citenamefont {Kavokin}\ \emph {et~al.}(2007)\citenamefont
  {Kavokin}, \citenamefont {Baumberg}, \citenamefont {Malpuech},\ and\
  \citenamefont {Laussy}}]{Kavokin2007}%
  \BibitemOpen
  \bibfield  {author} {\bibinfo {author} {\bibfnamefont {A.}~\bibnamefont
  {Kavokin}}, \bibinfo {author} {\bibfnamefont {J.~J.}\ \bibnamefont
  {Baumberg}}, \bibinfo {author} {\bibfnamefont {G.}~\bibnamefont {Malpuech}},
  \ and\ \bibinfo {author} {\bibfnamefont {F.~P.}\ \bibnamefont {Laussy}},\
  }\href@noop {} {\emph {\bibinfo {title} {{Microcavities}}}}\ (\bibinfo
  {publisher} {OUP Oxford},\ \bibinfo {year} {2007})\BibitemShut {NoStop}%
\bibitem [{\citenamefont {Lu}\ \emph {et~al.}(2016)\citenamefont {Lu},
  \citenamefont {Li}, \citenamefont {Hwang}, \citenamefont {Ofori-Okai},
  \citenamefont {Kurihara}, \citenamefont {Suemoto},\ and\ \citenamefont
  {Nelson}}]{Lu2016}%
  \BibitemOpen
  \bibfield  {author} {\bibinfo {author} {\bibfnamefont {J.}~\bibnamefont
  {Lu}}, \bibinfo {author} {\bibfnamefont {X.}~\bibnamefont {Li}}, \bibinfo
  {author} {\bibfnamefont {H.~Y.}\ \bibnamefont {Hwang}}, \bibinfo {author}
  {\bibfnamefont {B.~K.}\ \bibnamefont {Ofori-Okai}}, \bibinfo {author}
  {\bibfnamefont {T.}~\bibnamefont {Kurihara}}, \bibinfo {author}
  {\bibfnamefont {T.}~\bibnamefont {Suemoto}}, \ and\ \bibinfo {author}
  {\bibfnamefont {K.~A.}\ \bibnamefont {Nelson}},\ }\href@noop {} {\ ,\
  \bibinfo {pages} {Preprint at https://arxiv.org/abs/1605.06476} (\bibinfo
  {year} {2016})}\BibitemShut {NoStop}%
\bibitem [{\citenamefont {Baierl}\ \emph {et~al.}(2016)\citenamefont {Baierl},
  \citenamefont {Hohenleutner}, \citenamefont {Kampfrath}, \citenamefont
  {Zvezdin}, \citenamefont {Kimel}, \citenamefont {Huber},\ and\ \citenamefont
  {Mikhaylovskiy}}]{Baierl2016}%
  \BibitemOpen
  \bibfield  {author} {\bibinfo {author} {\bibfnamefont {S.}~\bibnamefont
  {Baierl}}, \bibinfo {author} {\bibfnamefont {M.}~\bibnamefont
  {Hohenleutner}}, \bibinfo {author} {\bibfnamefont {T.}~\bibnamefont
  {Kampfrath}}, \bibinfo {author} {\bibfnamefont {A.~K.}\ \bibnamefont
  {Zvezdin}}, \bibinfo {author} {\bibfnamefont {A.~V.}\ \bibnamefont {Kimel}},
  \bibinfo {author} {\bibfnamefont {R.}~\bibnamefont {Huber}}, \ and\ \bibinfo
  {author} {\bibfnamefont {R.~V.}\ \bibnamefont {Mikhaylovskiy}},\ }\href@noop
  {} {\bibfield  {journal} {\bibinfo  {journal} {Nature Photonics}\ }\textbf
  {\bibinfo {volume} {10}},\ \bibinfo {pages} {715} (\bibinfo {year}
  {2016})}\BibitemShut {NoStop}%
\bibitem [{\citenamefont {Amnon}\ and\ \citenamefont {Yeh}(2007)}]{Amnon2007}%
  \BibitemOpen
  \bibfield  {author} {\bibinfo {author} {\bibfnamefont {Y.}~\bibnamefont
  {Amnon}}\ and\ \bibinfo {author} {\bibfnamefont {P.}~\bibnamefont {Yeh}},\
  }\href@noop {} {\emph {\bibinfo {title} {{Optical Electronics in Modern
  Communications}}}},\ \bibinfo {edition} {6th}\ ed.\ (\bibinfo  {publisher}
  {Oxford University Press},\ \bibinfo {address} {Oxford, New York},\ \bibinfo
  {year} {2007})\ pp.\ \bibinfo {pages} {110--116}\BibitemShut {NoStop}%
\bibitem [{\citenamefont {Eddins}\ \emph {et~al.}(2013)\citenamefont {Eddins},
  \citenamefont {Beedle}, \citenamefont {Hendrickson},\ and\ \citenamefont
  {Friedman}}]{Eddins2013}%
  \BibitemOpen
  \bibfield  {author} {\bibinfo {author} {\bibfnamefont {A.~W.}\ \bibnamefont
  {Eddins}}, \bibinfo {author} {\bibfnamefont {C.~C.}\ \bibnamefont {Beedle}},
  \bibinfo {author} {\bibfnamefont {D.~N.}\ \bibnamefont {Hendrickson}}, \ and\
  \bibinfo {author} {\bibfnamefont {J.~R.}\ \bibnamefont {Friedman}},\ }\href
  {\doibase 10.1103/PhysRevLett.112.120501} {\bibfield  {journal} {\bibinfo
  {journal} {Physical Review Letters}\ }\textbf {\bibinfo {volume} {112}}
  (\bibinfo {year} {2013}),\ 10.1103/PhysRevLett.112.120501},\ \Eprint
  {http://arxiv.org/abs/1210.7330v1} {arXiv:1210.7330v1} \BibitemShut {NoStop}%
\bibitem [{\citenamefont {Zheng}\ and\ \citenamefont {Guo}(2000)}]{Zheng2000}%
  \BibitemOpen
  \bibfield  {author} {\bibinfo {author} {\bibfnamefont {S.~B.}\ \bibnamefont
  {Zheng}}\ and\ \bibinfo {author} {\bibfnamefont {G.~C.}\ \bibnamefont
  {Guo}},\ }\href {\doibase 10.1103/PhysRevLett.85.2392} {\bibfield  {journal}
  {\bibinfo  {journal} {Physical Review Letters}\ }\textbf {\bibinfo {volume}
  {85}},\ \bibinfo {pages} {2392} (\bibinfo {year} {2000})}\BibitemShut
  {NoStop}%
\bibitem [{\citenamefont {Pimenov}\ \emph {et~al.}(2006)\citenamefont
  {Pimenov}, \citenamefont {Mukhin}, \citenamefont {Ivanov}, \citenamefont
  {Travkin}, \citenamefont {Balbashov},\ and\ \citenamefont
  {Loidl}}]{Pimenov2006}%
  \BibitemOpen
  \bibfield  {author} {\bibinfo {author} {\bibfnamefont {A.}~\bibnamefont
  {Pimenov}}, \bibinfo {author} {\bibfnamefont {A.~A.}\ \bibnamefont {Mukhin}},
  \bibinfo {author} {\bibfnamefont {V.~Y.}\ \bibnamefont {Ivanov}}, \bibinfo
  {author} {\bibfnamefont {V.~D.}\ \bibnamefont {Travkin}}, \bibinfo {author}
  {\bibfnamefont {A.~M.}\ \bibnamefont {Balbashov}}, \ and\ \bibinfo {author}
  {\bibfnamefont {A.}~\bibnamefont {Loidl}},\ }\href {\doibase
  10.1038/nphys212} {\bibfield  {journal} {\bibinfo  {journal} {Nature
  Physics}\ }\textbf {\bibinfo {volume} {2}},\ \bibinfo {pages} {97} (\bibinfo
  {year} {2006})},\ \Eprint {http://arxiv.org/abs/0602173} {arXiv:0602173
  [cond-mat]} \BibitemShut {NoStop}%
\bibitem [{\citenamefont {Chumak}\ \emph {et~al.}(2015)\citenamefont {Chumak},
  \citenamefont {Vasyuchka}, \citenamefont {Serga},\ and\ \citenamefont
  {Hillebrands}}]{Chumak2015}%
  \BibitemOpen
  \bibfield  {author} {\bibinfo {author} {\bibfnamefont {A.~V.}\ \bibnamefont
  {Chumak}}, \bibinfo {author} {\bibfnamefont {V.~I.}\ \bibnamefont
  {Vasyuchka}}, \bibinfo {author} {\bibfnamefont {A.~A.}\ \bibnamefont
  {Serga}}, \ and\ \bibinfo {author} {\bibfnamefont {B.}~\bibnamefont
  {Hillebrands}},\ }\href@noop {} {\bibfield  {journal} {\bibinfo  {journal}
  {Nature Physics}\ }\textbf {\bibinfo {volume} {11}},\ \bibinfo {pages} {453}
  (\bibinfo {year} {2015})}\BibitemShut {NoStop}%
\bibitem [{\citenamefont {Kurebayashi}\ \emph {et~al.}(2011)\citenamefont
  {Kurebayashi}, \citenamefont {Dzyapko}, \citenamefont {Demidov},
  \citenamefont {Fang}, \citenamefont {Ferguson},\ and\ \citenamefont
  {Demokritov}}]{Kurebayashi2011}%
  \BibitemOpen
  \bibfield  {author} {\bibinfo {author} {\bibfnamefont {H.}~\bibnamefont
  {Kurebayashi}}, \bibinfo {author} {\bibfnamefont {O.}~\bibnamefont
  {Dzyapko}}, \bibinfo {author} {\bibfnamefont {V.~E.}\ \bibnamefont
  {Demidov}}, \bibinfo {author} {\bibfnamefont {D.}~\bibnamefont {Fang}},
  \bibinfo {author} {\bibfnamefont {A.~J.}\ \bibnamefont {Ferguson}}, \ and\
  \bibinfo {author} {\bibfnamefont {S.~O.}\ \bibnamefont {Demokritov}},\ }\href
  {\doibase 10.1038/nmat3053} {\bibfield  {journal} {\bibinfo  {journal}
  {Nature materials}\ }\textbf {\bibinfo {volume} {10}},\ \bibinfo {pages}
  {660} (\bibinfo {year} {2011})}\BibitemShut {NoStop}%
\bibitem [{\citenamefont {Hahn}\ \emph {et~al.}(2014)\citenamefont {Hahn},
  \citenamefont {Podlesnyak}, \citenamefont {Ehlers}, \citenamefont {Granroth},
  \citenamefont {Fishman}, \citenamefont {Kolesnikov}, \citenamefont
  {Pomjakushina},\ and\ \citenamefont {Conder}}]{Hahn2014}%
  \BibitemOpen
  \bibfield  {author} {\bibinfo {author} {\bibfnamefont {S.~E.}\ \bibnamefont
  {Hahn}}, \bibinfo {author} {\bibfnamefont {A.~A.}\ \bibnamefont
  {Podlesnyak}}, \bibinfo {author} {\bibfnamefont {G.}~\bibnamefont {Ehlers}},
  \bibinfo {author} {\bibfnamefont {G.~E.}\ \bibnamefont {Granroth}}, \bibinfo
  {author} {\bibfnamefont {R.~S.}\ \bibnamefont {Fishman}}, \bibinfo {author}
  {\bibfnamefont {A.~I.}\ \bibnamefont {Kolesnikov}}, \bibinfo {author}
  {\bibfnamefont {E.}~\bibnamefont {Pomjakushina}}, \ and\ \bibinfo {author}
  {\bibfnamefont {K.}~\bibnamefont {Conder}},\ }\href@noop {} {\bibfield
  {journal} {\bibinfo  {journal} {Physical Review B}\ }\textbf {\bibinfo
  {volume} {89}},\ \bibinfo {pages} {14420} (\bibinfo {year}
  {2014})}\BibitemShut {NoStop}%
\bibitem [{\citenamefont {Verd{\'{u}}}\ \emph {et~al.}(2009)\citenamefont
  {Verd{\'{u}}}, \citenamefont {Zoubi}, \citenamefont {Koller}, \citenamefont
  {Majer}, \citenamefont {Ritsch},\ and\ \citenamefont
  {Schmiedmayer}}]{Verd??2009}%
  \BibitemOpen
  \bibfield  {author} {\bibinfo {author} {\bibfnamefont {J.}~\bibnamefont
  {Verd{\'{u}}}}, \bibinfo {author} {\bibfnamefont {H.}~\bibnamefont {Zoubi}},
  \bibinfo {author} {\bibfnamefont {C.}~\bibnamefont {Koller}}, \bibinfo
  {author} {\bibfnamefont {J.}~\bibnamefont {Majer}}, \bibinfo {author}
  {\bibfnamefont {H.}~\bibnamefont {Ritsch}}, \ and\ \bibinfo {author}
  {\bibfnamefont {J.}~\bibnamefont {Schmiedmayer}},\ }\href {\doibase
  10.1103/PhysRevLett.103.043603} {\bibfield  {journal} {\bibinfo  {journal}
  {Physical Review Letters}\ }\textbf {\bibinfo {volume} {103}} (\bibinfo
  {year} {2009}),\ 10.1103/PhysRevLett.103.043603},\ \Eprint
  {http://arxiv.org/abs/0809.2552} {arXiv:0809.2552} \BibitemShut {NoStop}%
\bibitem [{\citenamefont {Spreeuw}\ \emph {et~al.}(1990)\citenamefont
  {Spreeuw}, \citenamefont {{Van Druten}}, \citenamefont {Beijersbergen},
  \citenamefont {Eliel},\ and\ \citenamefont {Woerdman}}]{Spreeuw1990}%
  \BibitemOpen
  \bibfield  {author} {\bibinfo {author} {\bibfnamefont {R.~J.~C.}\
  \bibnamefont {Spreeuw}}, \bibinfo {author} {\bibfnamefont {N.~J.}\
  \bibnamefont {{Van Druten}}}, \bibinfo {author} {\bibfnamefont {M.~W.}\
  \bibnamefont {Beijersbergen}}, \bibinfo {author} {\bibfnamefont {E.~R.}\
  \bibnamefont {Eliel}}, \ and\ \bibinfo {author} {\bibfnamefont {J.~P.}\
  \bibnamefont {Woerdman}},\ }\href {\doibase 10.1103/PhysRevLett.65.2642}
  {\bibfield  {journal} {\bibinfo  {journal} {Physical Review Letters}\
  }\textbf {\bibinfo {volume} {65}},\ \bibinfo {pages} {2642} (\bibinfo {year}
  {1990})}\BibitemShut {NoStop}%
\bibitem [{\citenamefont {Petit}\ \emph {et~al.}(2007)\citenamefont {Petit},
  \citenamefont {Moussa}, \citenamefont {Hennion}, \citenamefont
  {Pailh{\`{e}}s}, \citenamefont {Pinsard-Gaudart},\ and\ \citenamefont
  {Ivanov}}]{Petit2007}%
  \BibitemOpen
  \bibfield  {author} {\bibinfo {author} {\bibfnamefont {S.}~\bibnamefont
  {Petit}}, \bibinfo {author} {\bibfnamefont {F.}~\bibnamefont {Moussa}},
  \bibinfo {author} {\bibfnamefont {M.}~\bibnamefont {Hennion}}, \bibinfo
  {author} {\bibfnamefont {S.}~\bibnamefont {Pailh{\`{e}}s}}, \bibinfo {author}
  {\bibfnamefont {L.}~\bibnamefont {Pinsard-Gaudart}}, \ and\ \bibinfo {author}
  {\bibfnamefont {A.}~\bibnamefont {Ivanov}},\ }\href {\doibase
  10.1103/PhysRevLett.99.266604} {\bibfield  {journal} {\bibinfo  {journal}
  {Physical Review Letters}\ }\textbf {\bibinfo {volume} {99}},\ \bibinfo
  {pages} {1} (\bibinfo {year} {2007})}\BibitemShut {NoStop}%
\bibitem [{\citenamefont {Faucheaux}\ \emph {et~al.}(2014)\citenamefont
  {Faucheaux}, \citenamefont {Fu},\ and\ \citenamefont {Jain}}]{Faucheaux2014}%
  \BibitemOpen
  \bibfield  {author} {\bibinfo {author} {\bibfnamefont {J.~A.}\ \bibnamefont
  {Faucheaux}}, \bibinfo {author} {\bibfnamefont {J.}~\bibnamefont {Fu}}, \
  and\ \bibinfo {author} {\bibfnamefont {P.~K.}\ \bibnamefont {Jain}},\ }\href
  {\doibase 10.1021/jp412157c} {\bibfield  {journal} {\bibinfo  {journal}
  {Journal of Physical Chemistry C}\ }\textbf {\bibinfo {volume} {118}},\
  \bibinfo {pages} {2710} (\bibinfo {year} {2014})}\BibitemShut {NoStop}%
\bibitem [{\citenamefont {Yamaguchi}\ \emph {et~al.}(2013)\citenamefont
  {Yamaguchi}, \citenamefont {Kurihara}, \citenamefont {Minami}, \citenamefont
  {Nakajima},\ and\ \citenamefont {Suemoto}}]{Yamaguchi2013}%
  \BibitemOpen
  \bibfield  {author} {\bibinfo {author} {\bibfnamefont {K.}~\bibnamefont
  {Yamaguchi}}, \bibinfo {author} {\bibfnamefont {T.}~\bibnamefont {Kurihara}},
  \bibinfo {author} {\bibfnamefont {Y.}~\bibnamefont {Minami}}, \bibinfo
  {author} {\bibfnamefont {M.}~\bibnamefont {Nakajima}}, \ and\ \bibinfo
  {author} {\bibfnamefont {T.}~\bibnamefont {Suemoto}},\ }\href {\doibase
  10.1103/PhysRevLett.110.137204} {\bibfield  {journal} {\bibinfo  {journal}
  {Physical Review Letters}\ }\textbf {\bibinfo {volume} {110}},\ \bibinfo
  {pages} {137204} (\bibinfo {year} {2013})}\BibitemShut {NoStop}%
\bibitem [{\citenamefont {Kimel}\ \emph {et~al.}(2005)\citenamefont {Kimel},
  \citenamefont {Kirilyuk}, \citenamefont {Usachev}, \citenamefont {Pisarev},
  \citenamefont {Balbashov},\ and\ \citenamefont {Rasing}}]{Kimel2005}%
  \BibitemOpen
  \bibfield  {author} {\bibinfo {author} {\bibfnamefont {a.~V.}\ \bibnamefont
  {Kimel}}, \bibinfo {author} {\bibfnamefont {A.}~\bibnamefont {Kirilyuk}},
  \bibinfo {author} {\bibfnamefont {P.~a.}\ \bibnamefont {Usachev}}, \bibinfo
  {author} {\bibfnamefont {R.~V.}\ \bibnamefont {Pisarev}}, \bibinfo {author}
  {\bibfnamefont {a.~M.}\ \bibnamefont {Balbashov}}, \ and\ \bibinfo {author}
  {\bibfnamefont {T.}~\bibnamefont {Rasing}},\ }\href@noop {} {\bibfield
  {journal} {\bibinfo  {journal} {Nature}\ }\textbf {\bibinfo {volume} {435}},\
  \bibinfo {pages} {655} (\bibinfo {year} {2005})}\BibitemShut {NoStop}%
\bibitem [{\citenamefont {Koshizuka}\ and\ \citenamefont
  {Hayashi}(1988)}]{Koshizuka1988}%
  \BibitemOpen
  \bibfield  {author} {\bibinfo {author} {\bibfnamefont {N.}~\bibnamefont
  {Koshizuka}}\ and\ \bibinfo {author} {\bibfnamefont {K.}~\bibnamefont
  {Hayashi}},\ }\href@noop {} {\bibfield  {journal} {\bibinfo  {journal}
  {Journal of the Physical Society of Japan}\ }\textbf {\bibinfo {volume}
  {57}},\ \bibinfo {pages} {4418} (\bibinfo {year} {1988})}\BibitemShut
  {NoStop}%
\bibitem [{\citenamefont {Yamada}\ \emph {et~al.}(1996)\citenamefont {Yamada},
  \citenamefont {Saitoh},\ and\ \citenamefont {Ooki}}]{Yamada1996}%
  \BibitemOpen
  \bibfield  {author} {\bibinfo {author} {\bibfnamefont {M.}~\bibnamefont
  {Yamada}}, \bibinfo {author} {\bibfnamefont {M.}~\bibnamefont {Saitoh}}, \
  and\ \bibinfo {author} {\bibfnamefont {H.}~\bibnamefont {Ooki}},\ }\href
  {\doibase 10.1063/1.117015} {\bibfield  {journal} {\bibinfo  {journal}
  {Applied Physics Letters}\ }\textbf {\bibinfo {volume} {69}},\ \bibinfo
  {pages} {3659} (\bibinfo {year} {1996})}\BibitemShut {NoStop}%
\bibitem [{\citenamefont {Serga}\ \emph {et~al.}(2010)\citenamefont {Serga},
  \citenamefont {Chumak},\ and\ \citenamefont {Hillebrands}}]{Serga2010}%
  \BibitemOpen
  \bibfield  {author} {\bibinfo {author} {\bibfnamefont {a.~a.}\ \bibnamefont
  {Serga}}, \bibinfo {author} {\bibfnamefont {a.~V.}\ \bibnamefont {Chumak}}, \
  and\ \bibinfo {author} {\bibfnamefont {B.}~\bibnamefont {Hillebrands}},\
  }\href {\doibase 10.1088/0022-3727/43/26/264002} {\bibfield  {journal}
  {\bibinfo  {journal} {Journal of Physics D: Applied Physics}\ }\textbf
  {\bibinfo {volume} {43}},\ \bibinfo {pages} {264002} (\bibinfo {year}
  {2010})}\BibitemShut {NoStop}%
\bibitem [{\citenamefont {Lin}\ \emph {et~al.}(2009)\citenamefont {Lin},
  \citenamefont {Werley},\ and\ \citenamefont {Nelson}}]{Lin2009a}%
  \BibitemOpen
  \bibfield  {author} {\bibinfo {author} {\bibfnamefont {K.-H.}\ \bibnamefont
  {Lin}}, \bibinfo {author} {\bibfnamefont {C.~A.}\ \bibnamefont {Werley}}, \
  and\ \bibinfo {author} {\bibfnamefont {K.~A.}\ \bibnamefont {Nelson}},\
  }\href@noop {} {\bibfield  {journal} {\bibinfo  {journal} {Appl. Phys.
  Lett.}\ }\textbf {\bibinfo {volume} {95}},\ \bibinfo {pages} {103304}
  (\bibinfo {year} {2009})}\BibitemShut {NoStop}%
\bibitem [{\citenamefont {Reid}\ \emph {et~al.}(2015)\citenamefont {Reid},
  \citenamefont {Rasing}, \citenamefont {Pisarev}, \citenamefont {D{\"{u}}rr},\
  and\ \citenamefont {Hoffmann}}]{Reid2015}%
  \BibitemOpen
  \bibfield  {author} {\bibinfo {author} {\bibfnamefont {A.~H.~M.}\
  \bibnamefont {Reid}}, \bibinfo {author} {\bibfnamefont {T.}~\bibnamefont
  {Rasing}}, \bibinfo {author} {\bibfnamefont {R.~V.}\ \bibnamefont {Pisarev}},
  \bibinfo {author} {\bibfnamefont {H.~A.}\ \bibnamefont {D{\"{u}}rr}}, \ and\
  \bibinfo {author} {\bibfnamefont {M.~C.}\ \bibnamefont {Hoffmann}},\ }\href
  {\doibase 10.1063/1.4908186} {\bibfield  {journal} {\bibinfo  {journal}
  {Applied Physics Letters}\ }\textbf {\bibinfo {volume} {106}} (\bibinfo
  {year} {2015}),\ 10.1063/1.4908186}\BibitemShut {NoStop}%
\bibitem [{\citenamefont {Sivarajah}\ \emph {et~al.}(2015)\citenamefont
  {Sivarajah}, \citenamefont {Ofori-Okai}, \citenamefont {Teo}, \citenamefont
  {Werley},\ and\ \citenamefont {Nelson}}]{Sivarajah2015}%
  \BibitemOpen
  \bibfield  {author} {\bibinfo {author} {\bibfnamefont {P.}~\bibnamefont
  {Sivarajah}}, \bibinfo {author} {\bibfnamefont {B.~K.}\ \bibnamefont
  {Ofori-Okai}}, \bibinfo {author} {\bibfnamefont {S.~M.}\ \bibnamefont {Teo}},
  \bibinfo {author} {\bibfnamefont {C.~A.}\ \bibnamefont {Werley}}, \ and\
  \bibinfo {author} {\bibfnamefont {K.~A.}\ \bibnamefont {Nelson}},\
  }\href@noop {} {\bibfield  {journal} {\bibinfo  {journal} {New Journal of
  Physics}\ }\textbf {\bibinfo {volume} {17}} (\bibinfo {year}
  {2015})}\BibitemShut {NoStop}%
\bibitem [{\citenamefont {Wallraff}\ \emph
  {et~al.}(2004{\natexlab{b}})\citenamefont {Wallraff}, \citenamefont
  {Schuster}, \citenamefont {Blais},\ and\ \citenamefont
  {Frunzio}}]{Wallraff2004a}%
  \BibitemOpen
  \bibfield  {author} {\bibinfo {author} {\bibfnamefont {A.}~\bibnamefont
  {Wallraff}}, \bibinfo {author} {\bibfnamefont {D.}~\bibnamefont {Schuster}},
  \bibinfo {author} {\bibfnamefont {A.}~\bibnamefont {Blais}}, \ and\ \bibinfo
  {author} {\bibfnamefont {L.}~\bibnamefont {Frunzio}},\ }\href {\doibase
  10.1038/nature02851} {\bibfield  {journal} {\bibinfo  {journal} {Nature}\
  }\textbf {\bibinfo {volume} {431}},\ \bibinfo {pages} {162} (\bibinfo {year}
  {2004}{\natexlab{b}})},\ \Eprint {http://arxiv.org/abs/0407325}
  {arXiv:0407325 [cond-mat]} \BibitemShut {NoStop}%
\end{thebibliography}

\end{document}